\documentclass[fleqn,usenatbib]{mnras}

\usepackage{graphicx}	% Including figure files
\usepackage{amsmath}	% Advanced maths commands
\usepackage{amssymb}	% Extra maths symbols
\usepackage{multicol}        % Multi-column entries in tables
\usepackage{bm}		% Bold maths symbols, including upright Greek
\usepackage{pdflscape}	% Landscape pages
\usepackage{hyperref}
\usepackage{amsfonts}
\usepackage{txfonts}
\usepackage{ctable}
\usepackage{url}
\usepackage[normalem]{ulem}

\usepackage{color}
\usepackage{grffile}

\newcommand\altaffilmark[1]{$^{#1}$}
\newcommand\altaffiltext[1]{$^{#1}$}
\newcommand{\Ocool}{\Lambda_{\rm O}}
\newcommand{\Cp}{{\rm C}^{+}}
\newcommand{\Cpcool}{\Lambda_{\Cp}}

\newcommand{\driftvel}{{\bm v}_d}

\newcommand{\TAD}{T_{+{\rm AD}}}

\newcommand \bB         {{\bf B}}

\newcommand \beam {{\sigma}}
\newcommand \bbhat      {\hat{\bf b}}
\newcommand \Cbar {\bar{C}}
\newcommand \Cpol {C_{\rm pol}}

\newcommand \bE         {{\bf E}}

\newcommand \falign {f_{\rm align}}

\newcommand \bahat {\hat{\bf a} }
\newcommand \bxhat      {\hat{\bf x}}
\newcommand \byhat      {\hat{\bf y}}

\newcommand \beq        {\begin{equation}}
\newcommand \beqa	{\begin{eqnarray}}

\newcommand \cm         {\,{\rm cm}}

\newcommand \eeq	{\end{equation}}
\newcommand \eeqa	{\end{eqnarray}}
\newcommand \erg	{\,{\rm ergs}}

\newcommand \gtsim	{\gtrsim}		 %apj version
\newcommand \Ha 	{{\rm H}}
\newcommand \He	        {{\rm He}}
\newcommand \HH	        {{\rm H}_2}

\newcommand \K  	{\,{\rm K}}

\newcommand \ltsim	{\lesssim}		 %apj version

\newcommand \muG        {\mu{\rm G}}

\newcommand \nH         {n_{\rm H}}
\newcommand \pc  	{\,{\rm pc}}

\newcommand \s	        {\,{\rm s}}

\newcommand \Teff       {T_{\rm eff}}

\newcommand \xtimes     {{\!\times\!}}
\newcommand \chemphys {{Chem.~Phys.}}

\newcommand{\oldtext}[1]{}

\newcommand{\newtext}[1]{{ #1}}

\usepackage[T1]{fontenc}
\usepackage{ae,aecompl}

\title[CNM Turbulence]{Turbulent dissipation, CH$^+$ abundance, $\HH$ line luminosities, and polarization in the cold neutral medium
 \vspace{-0.5cm}}

\author[Moseley et al.]{
\parbox[t]{\textwidth}{ 
	Eric R.~Moseley\altaffilmark{1}\thanks{E-mail: moseley@princeton.edu},
	B.~T.~Draine\altaffilmark{1},
	Kengo Tomida\altaffilmark{2},
	James M. Stone\altaffilmark{1,3}
} 
\vspace*{6pt} \\
\altaffiltext{1}{Department of Astrophysical Sciences, Princeton University, Princeton, NJ 08540, USA}\\
\altaffiltext{2}{Astronomical Institute, Tohoku University, Sendai, Miyagi 980-8578, Japan
}\\
\altaffiltext{3}{School of Natural Sciences, Institute for Advanced Study, Princeton, NJ 08544, USA
}
}

\date{}

\pubyear{2020}

\begin{document}
%\label{firstpage}
\maketitle

\begin{abstract}  In the cold neutral medium, high out-of-equilibrium temperatures are created by intermittent dissipation processes, including shocks, viscous heating, and ambipolar diffusion.
The 
high-temperature excursions are thought to explain the enhanced abundance of CH$^{+}$ observed along diffuse molecular sight-lines. Intermittent high temperatures should also have an impact on H$_2$ line luminosities. We carry out simulations of MHD turbulence in molecular clouds including heating and cooling, and post-process them to study $\HH$ line emission and hot-gas chemistry, particularly the formation of CH$^+$. We explore multiple magnetic field strengths and equations of state. We use a new $\HH$ cooling function for $\nH \leq 10^5\cm^{-3}$, $T\leq 5000\K$, and variable $\HH$ fraction. We make two important simplifying assumptions: (i) the $\HH/{\rm H}$ fraction is fixed everywhere, and (ii) we exclude from our analysis regions where the ion-neutral drift velocity is calculated to be greater than 5 km/s. Our models produce $\HH$ emission lines in accord with many observations, although extra excitation mechanisms are required in some clouds. For realistic r.m.s. magnetic field strengths ($\approx 10$ $\mu$G) and velocity dispersions, we reproduce observed CH$^+$ abundances. These findings contrast with those of Valdivia et al. (2017).
Comparison of predicted dust polarization with observations by {\it Planck} suggests that the mean field $\gtsim 5\muG$, so that the turbulence is sub-Alfv\'enic.
We recommend future work treating ions and neutrals as separate fluids to more accurately capture the effects of ambipolar diffusion on CH$^+$ abundance.
\end{abstract}


%
%%%%%%%%%%%%%%%%%%%%%%%%%%%%%%%%%%%%%%%%%%%%%%%%%
\begin{figure*}
\begin{center}
	\includegraphics[width=0.8\textwidth]{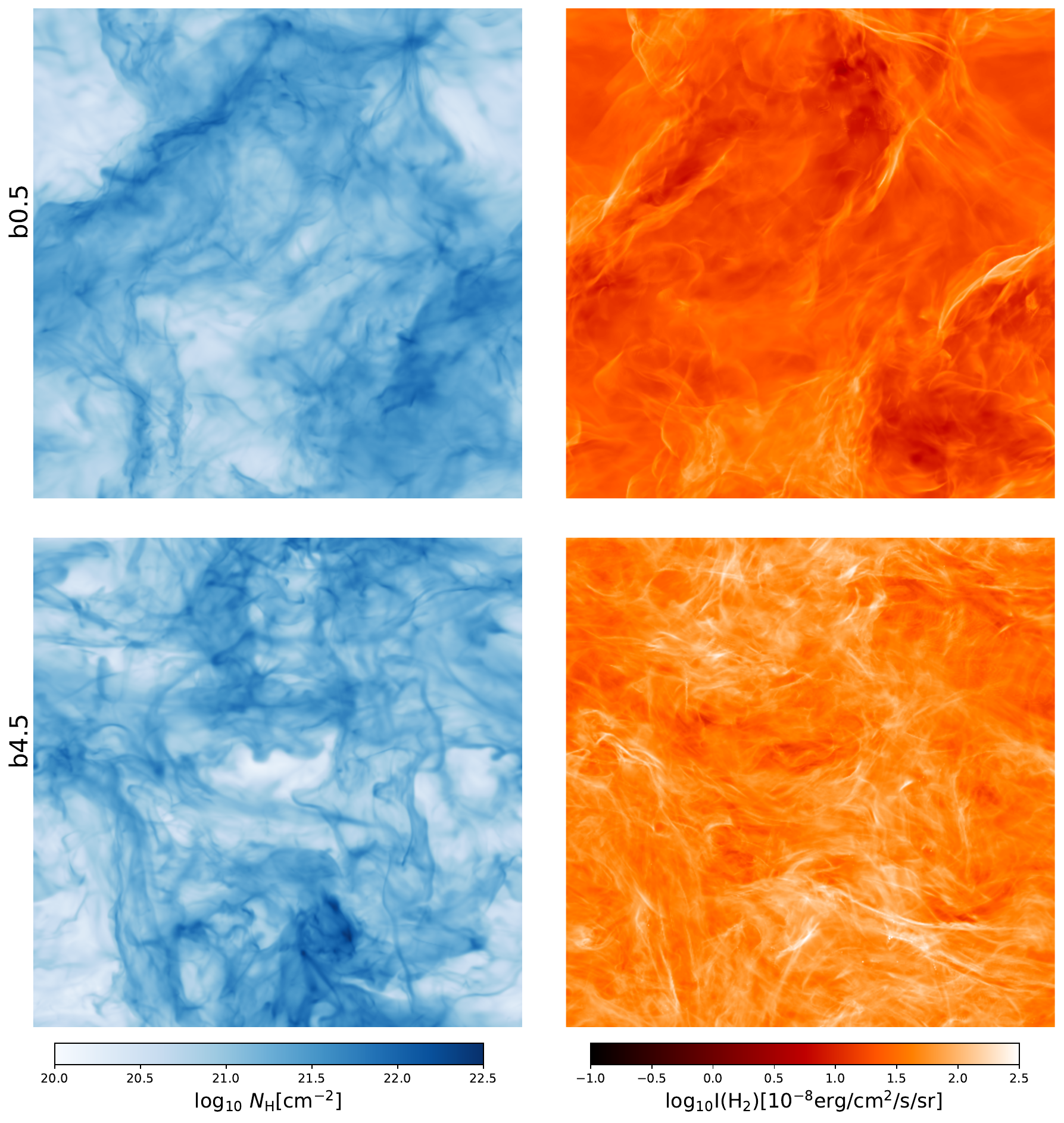}
    \caption{{\bf Left:} Column density of H nuclei in a snapshot of each of our $512^3$ ideal MHD simulation including heating and cooling processes.
    {\bf Right:} Total $\HH$ rotational line emission intensity. It is apparent that much of the emission comes from shocks by the sheet-like distribution of the emission (See Sec.~\ref{sec:lineemission}). For each simulation, the box is 20 pc on a side with a mean density of $n_{H} = 30$ cm$^{-3}$. The
   initial (and therefore mean) vector magnetic field 
    $B_0 = 0.5\muG$ (top) and $4.5\muG$ (bottom), oriented in the horizontal direction.}
    \label{fig:column_density}
\end{center}
\end{figure*}
%%%%%%%%%%%%%%%%%%%%%%%%%%%%%%%%%%%%%%%%%%%%%%%%%
%
%
%%%%%%%%%%%%%%%%%%%%%%%%%%%%%%%%%%%%%%%%%%%%%%%%%%

%%%%%%%%%%%%%%%%% BODY OF PAPER %%%%%%%%%%%%%%%%%%
\section{Introduction}\label{sec:intro}
The abundance of CH$^+$ in diffuse interstellar clouds has been a challenge to explain since it was first identified \citep{Douglas+Herzberg_1941}. The observed column densities \citep[$\gtrsim 10^{13} \,{\rm cm}^{-2}$,][]{LambertDanks1986ApJ} are puzzling due to the multiple efficient destruction mechanisms for CH$^+$: reactions with H, $\HH$, $e^-$, and dissociation by ultraviolet radiation. In addition, the reaction
\begin{equation}
    {\rm C}^+ + \HH \rightarrow {\rm CH}^+ + {\rm H} \hspace{1.0cm} \Delta E/k = 4640\hspace{0.1cm}{\rm K},\label{eq:react}
\end{equation}
is strongly endothermic and only proceeds appreciably for temperatures $T\gtrsim 1000\K$, significantly higher than the $\approx 100\K$ temperatures characteristic of these clouds. As a result, steady state models fail to produce CH$^+$ in sufficient quantities, predicting column densities at least two orders magnitude below observed values \citep{van_Dishoeck+Black_1986}. 

Proposed solutions to this problem rely on ways to heat some fraction of the gas, even transiently, to $T\gtrsim 1000\K$.
\citet{Elitzur+Watson_1978} proposed that shock waves in diffuse molecular clouds could account for the CH$^+$ production, and two-fluid MHD shock models were subsequently used to model CH$^+$ formation
\citep{Flower+Pineau_des_Foret+Hartquist_1985,Draine+Katz_1986a,Draine+Katz_1986b,Draine_1986b}.
Other solutions have also been proposed, including diffuse gas undergoing strong photoelectric heating \citep{white1984interstellar},
dense photon-dominated regions (PDRs) \citep{duley1992ch+, sternberg1995apjs}, and heating in boundary layers at cloud surfaces \citep{duley1992ch+}. 

As diffuse interstellar clouds are supersonically turbulent, intermittent shock heating is one possible way to heat enough of the gas to these temperatures.
On the basis of laboratory experiments of turbulent flows, \citet{falgarone1995intermittency}  suggested that intermittent dissipation of turbulence  could heat diffuse molecular clouds to these temperatures.
\citet{pan2009temperature} found that, in compressible MHD turbulence simulations, a few percent of the gas by mass could be heated to $\gtrsim 1000\K$ in such diffuse molecular clouds, and thus produce the observed amounts of CH$^+$.  These findings may also help to explain observed levels of $\HH$ rotational line emission \citep[e.g.][]{Ingalls+Bania+Boulanger+etal_2011} as well. High-$j$ states of $\HH$ can only be effectively populated in relatively hot gas $\gtrsim 1000\K$ or through ultraviolet pumping. It is thus not surprising that CH$^+$ column density and
rotationally-excited $\HH$ are correlated \citep{FrischJura1980ApJ,LambertDanks1986ApJ}.

Drift between ionic and neutral species 
in MHD shocks has been proposed as a way to help overcome the energy barrier in reaction (\ref{eq:react}) \citep{Draine_1980, Flower+Pineau_des_Foret+Hartquist_1985}. 
\citet[][hereafter MML15]{Myers+McKee+Li_2015} 
%ran 
analyzed
MHD turbulence simulations with an isothermal equation of state for the purpose of addressing the CH$^+$ abundance and found that the contribution to the reaction 
rate from ion-neutral drift was the dominant effect responsible for generating CH$^+$. 

\newtext{In disagreement with the results of MML15 are those of \citet[][hereafter V17]{Valdivia+Godard+Hennebelle+etal_2017}. V17 ran a two-phase, colliding flow MHD simulation and found that CH$^+$ was \textit{not} primarily produced by high ion-neutral drift velocities in their simulations. Their simulations also under-produced CH$^+$ relative to observations, for reasons that they discuss. For one, it seems that the ion-neutral drift velocity distribution is not converged, and these velocities increase in magnitude at higher resolution. Second, due to the nature of colliding flow simulations that inject a single, warm phase of ISM into the box, they likely underestimate the $\HH$ fraction in low density regions. As well, highly excited $\HH$ may help to overcome the high reaction barrier present in reaction \ref{eq:react}.}
%\omittext{MHD simulations are likely the most realistic approach for understanding the full problem of CH$^+$ production. It allows for the characteristics of turbulent dissipation to be properly captured, and an informed estimate for the distribution of ion-neutral drift velocities can also be made. By assuming a balance between heating (cosmic ray ionizations, photoelectric emission from dust grains, and viscous dissipation) and cooling (C$^+$, O, and $\HH$ line emission), they determined a temperature in each cell of their simulation after the simulation had been completed. They then assumed an equilibrium ion-neutral drift velocity in each cell as a balance between the electromotive force and drag. In this way, they were able to estimate the CH$^+$ abundance and $\HH$ level populations in their simulation by scaling to physical units. }

\newtext{Our approach bears more similarity to that of MML15.} Like MML15, we have run ideal MHD turbulence simulations to understand the abundance of CH$^+$. We use the same abundances of H, $\HH$, $e^-$, C, and O, the same mean density, and the same box size to make our simulations as directly comparable to theirs as possible. Two of our simulations use a low initial magnetic field strength (0.5 $\mu$G), and two use a high magnetic field strength (4.5 $\mu$G). Two are run with an isothermal equation of state as in MML15, and two are run with an adiabatic ($\gamma = 5/3$) equation of state with heating and cooling processes included throughout the simulation. For all of our simulations, we also estimate the ambipolar diffusion heating in each cell with a post-processing scheme described in section~\ref{sec:adheat}. This effect is separate from the 
streaming-induced enhancement of the reaction rate,
and similarly important. In our simulations, the ambipolar diffusion heating rate
can become the dominant heating term in the low density regions where MML15 determined the majority of CH$^+$ is produced. We present maps of 
total
column density and $\HH$ rotational line intensities for those simulations that include heating and cooling processes throughout in Fig.~\ref{fig:column_density}.

Because turbulence leads to disorder in the magnetic field, we check to see whether the proposed levels of MHD turbulence are consistent with observations of polarized emission from aligned dust grains \citep{Planck_int_results_xxi_2015,Planck_2018_XII}.

In section~\ref{sec:methods} we review our
model, the effects of ambipolar diffusion, and the relevant heating processes. In section~\ref{sec:cool}, we detail our (new) calculation for the $\HH$ cooling function and an accurate fit to it for computational ease (see also App.~\ref{app:hh}), as well as describe the C$^+$ and O line cooling that we use. In section~\ref{sec:lineemission}, we discuss details of $\HH$ line emission. Then, in section~\ref{sec:chplus} we describe the chemistry that goes into producing CH$^+$.
We present our results for the temperature and drift velocity in our simulations, the CH$^+$ abundance, the velocity distributions of CH$^+$ molecules, $\HH$ rotational line emission, and the polarization of dust emission in section~\ref{sec:results}. Finally, we discuss our results in section~\ref{sec:discussion} and provide a summary of our findings in section~\ref{sec:summary}. A summary of our simulation parameters and results can be found in tables~\ref{tab:parameters} and~\ref{tab:sims}, respectively.

\section{Fluid dynamics}\label{sec:methods}

As described in Sec.~\ref{sec:intro}, our simulations are designed to study the formation of the CH$^+$ molecule and emission from $\HH$ rotational transitions in turbulent molecular clouds.

To investigate the importance of magnetic field strength, and to compare different treatments of the fluid dynamics, we have run a number of simulations.
In Appendix~\ref{app:cutoff}, we explore the numerical convergence of our results using resolutions ranging from $64^3$ through $512^3$. In the body of this paper however, we will focus on simulations run at a resolution of $512^3$.

For each magnetic field strength, we calculate the fluid motions assuming ideal MHD and explicit heating and cooling.  We also carry out simulations using ideal MHD and an isothermal equation of state, to evaluate the effects of heating and cooling on the fluid motions.
We post-process these isothermal simulations following a procedure similar to that given in MML15 (see Sec.~\ref{sec:heat}). 
Our post-processing of the isothermal simulations differs from MML15 in that we use
a new $\HH$ cooling function (Eq.~\ref{eq:H2cool}), and we attempt to estimate the effects of ambipolar diffusion heating on the system as described in the coming sections.

\subsection{Models \& Scaling}\label{sec:scaling}

Intermittent dissipation events in molecular clouds arise from the supersonic MHD turbulence that pervades them. 
To model these clouds, we use the results of four 512$^3$ driven MHD turbulence simulations run with the astrophysical MHD code Athena++.\footnote{https://github.com/PrincetonUniversity/athena-public-version} These simulations utilize periodic boundary conditions and a cubic domain. They are driven solenoidally between wavenumbers $k = 2\pi/\ell_0, 4\pi/\ell_0$ with a power spectrum $P(k) \propto k^{-2}$, where $\ell_0 = 20$ pc is the length of a side of the simulation volume. The driving follows the Ornstein-Uhlenbeck process \citep{lynn2012resonance}, smoothly evolving the driving with a correlation time of about 1/10 of a dynamical time. We use the Harten-Lax-van Leer-Discontinuities (HLLD) Riemann solver together with a second order Piecewise-Linear-Mesh (PLM) primitive reconstruction with a second order van Leer time integrator. 
As magnetic fields and their effects in molecular clouds may vary, we use two 
initial field strengths to compare to one another: 0.5 $\mu$G and 4.5 $\mu$G. 

Simulations begin with a uniform medium and uniform magnetic field ${\bm B}_0$, and are then driven until the velocity dispersion $\sigma_{\rm 3D}$ and the root-mean-square magnetic field strength $B_{\rm rms}$ saturate. We adjust the driving so that the saturated value of $\sigma_{\rm 3D}$ is close to the observed size-linewidth relation:

\beq
\sigma_{\rm 3D} \approx \sqrt{3}\,(0.72\,{\rm km/s})\,\bigg(\frac{R}{1 {\rm pc}}\bigg)^{0.5}
\approx 3.94 \,{\rm km/s},
\eeq
where we have let the cloud's radius $R = \ell_0/2 = 10$ pc. The driving power necessary to reach this will vary depending on field strength and equation of state. The values we have adopted for the size-linewidth relation are from
\citet{Solomon+Rivolo+Barrett+Yahil_1987}. 

The parameters we use are shown in Table~\ref{tab:parameters}. Many of these are chosen to reflect values from MML15. We define the plasma
$\beta$ as the ratio of the volume-averaged thermal pressure to the volume averaged magnetic pressure. The Mach number is the mass-weighted root-mean-square (rms) velocity divided by the mass-weighted sound speed, and the Alfv\'{e}n Mach number is the mass-weighted rms velocity divided by the volume averaged Alfv\'{e}n speed. These definitions are chosen so that the Mach number and Alfv\'en Mach number are simply related to the ratio of kinetic energy to thermal energy and kinetic energy to magnetic energy, respectively. Like MML15, we adopt a constant composition and assume all of our ions are C$^+$. 
%%%%%%%%%%%%%%%%%%%%%%%%%%%%%%%%%%%%%%%%%%%%%%%%%%
\begin{table}
    \centering
    \begin{tabular}{||c|c||}
    \hline
         $x({\rm H})\equiv n({\rm H})/n_{\rm H}$ & 0.68 \\
         $x(\HH)\equiv n(\HH)/n_{\rm H}$ & 0.16 \\
         $x({\rm He})\equiv n({\rm He})/n_{\rm H}$ & 0.1 \\
         $x(e^{-})\equiv n_e/n_{\rm H}$ & $1.6\times10^{-4}$ \\
         $x({\rm C})\equiv n({\rm C}^+)/n_{\rm H}$ &  $1.6\times10^{-4}$\\
         $x({\rm O})\equiv n({\rm O})/n_{\rm H}$ &  $3.2\times10^{-4}$\\
         $\mu$ & 1.49$m_{\rm proton}$\\
         $\langle n_{\rm H}\rangle$ & 30 cm$^{-3}$ \\
         $\langle \rho \rangle $ = $\mu \langle n \rangle $ & $7.0 \times10^{-23}$ g/cm$^{-3}$\\
         $\langle N_{\rm H}\rangle=\langle n_{\rm H}\rangle\ell_0$ & $1.85\times10^{21}$ cm$^{-2}$\\
         %\omittext{$\ell_0$} & \omittext{20 pc}\\
         %\omittext{$c_0$} & \omittext{0.38 km/s}\\
    \hline
    \end{tabular}
    \caption{Model Parameters. 
    %inputs of our model.
    }
    \label{tab:parameters}
\end{table}
%%%%%%%%%%%%%%%%%%%%%%%%%%%%%%%%%%%%%%%%%%%%%%%%%%

We 
%also 
neglect gravity.  The total mass in our volume is about 8300 M$_\odot$, giving an overall virial parameter $\alpha_{\rm vir} = 5 \sigma_{\rm 1D}^2 R/GM \approx 7.2$, rendering self-gravity negligible. We may also compare the effects of gravity to magnetic fields through the mass-to-flux ratio relative to critical
\begin{equation}
    \mu_{\Phi} = \frac{2\pi M\sqrt{G}}{B_{\rm rms} \ell_0^2}.
\end{equation}
For our simulations $\mu_{\Phi}$ ranges from about 0.7 to 1.9, so these simulations range from somewhat magnetically sub-critical to somewhat super-critical. 

We explicitly follow internal energy in our simulations and changes therein due to heating (from cosmic rays and photoelectric emission from dust grains) and cooling (due to C$^+$, O, and $\HH$ line emission). The cooling is not scale-free, introducing a particular length, time, and temperature. When using this cooling function as an explicit source term in a simulation, we are thus given less freedom than when post-processing isothermal simulations.

\subsection{Ambipolar Diffusion } \label{sec:AD}

\setlength{\tabcolsep}{3pt}
%
%%%%
%%%% A table of the simulations %%%%%%%%%%%%%%%%%%%%%%%%%%%%%%%%%%%%%%%%%%%%%%%%%%%%%%%%%%%%
\begin{table*}
\begin{center}
\begin{tabular}{|| c c c c c c c c c c c c c c||} 
\hline
\text{name} & $B_0$ & $B_{||,{\rm rms}}$ & $B_{\bot,{\rm rms}}/\sqrt{2}$ & $B_{\rm rms}$ & $\beta_0$& $\beta$ & $\sigma_{\rm 3D}$ & $\mathcal{M}_A$  &$\dot{\varepsilon}$ & $\log N_{{\rm CH}^+,50}$ & $I_{{\rm H}_2,50}$ & $\langle \tilde{P}_x\rangle$ & $\langle \tilde{P}_z\rangle$\\ %&  $\sigma_{{\rm CH}^+,50}$(km/s) & $\sigma_{{\rm H}_2,50}$(km/s)\\
\text{units} & $\mu$G &  $\mu$G &  $\mu$G &  $\mu$G & \text{None} & \text{None} & \text{km/s} & \text{None} & $10^{-26}$erg cm$^{-3}$ s$^{-1}$ & $\log$[cm$^{-2}$] & $10^{-8}$erg/s/cm$^{2}$/sr & \text{None} & \text{None}\\
\hline
b0.5 & 0.5 & 2.4 & 2.5 & 4.3 & 24 & 0.51 & 4.3 & 2.9 & 7.4 & 12.2 & 20.8 & 0.22 & 0.22 \\
b0.5-iso & 0.5 & 2.0 & 2.4 &  3.7 & 24 & 0.68 & 4.0 & 3.1 & 6.7 & 12.3& 20.4 & 0.18 & 0.21 \\
b4.5 & 4.5 & 6.1 & 5.1 & 9.5 & 0.38 & 0.10 & 3.9 & 1.2 & 10.0 & 13.2 & 45.9 & 0.27 & 0.30 \\
b4.5-iso & 4.5 & 5.8 & 4.4 & 8.6 & 0.38 & 0.13 & 4.2 & 1.4 & 6.7 & 13.1 & 47.8 & 0.24 & 0.29 \\
\hline
\end{tabular}
\end{center}
\caption{
Simulation parameters and selected statistics. Simulations labeled with ``iso'' 
employed an isothermal equation of state and are processed differently, as described in section \ref{sec:heat}. 
$B_0$ and $B_{\rm rms}$ are the mean and root-mean-square magnetic field strength; $B_{||,{\rm rms}}$ and $B_{\bot,{\rm rms}}/\sqrt{2}$ are the root-mean-square components of the saturated magnetic field along and across the mean magnetic field, respectively; $\beta_0$ and $\beta$ are the initial and final plasma beta; $\sigma_{3{\rm D}}$ is the 3D velocity dispersion; $\mathcal{M}_A$ is the final Alfv{\'e}n Mach number; $\dot{\epsilon}$ is the input driving power/volume; $N_{{\rm CH}^+,50}$ is the median CH$^+$ column density; $I_{{\rm H}_2,50}$ is the median sum total intensity of the $\HH$ rotational lines;
and $\langle \tilde{P}_x\rangle$ and $\langle{P}_z\rangle$ are the mean magnetic field alignment parameters for viewing along and across the mean magnetic field direction
(see Sec.~\ref{sec:polarization}).
}
\label{tab:sims}
\end{table*}

%%%%%%%%%%%%%%%%%%%%%%%%%%%%%%%%%%%%%%%%%%%%%%%%%%%%%%%%%%%%
%%%%
%%%%
%

The effects of ambipolar diffusion in the CNM have been treated in three separate ways in MHD turbulence simulations. %Need citations.
The first is to treat ions and neutrals as separate fluids that interact through a drag force
with frictional heating \citep{Draine_1986a,li2008sub}. This is the most true-to-life of the three methods, but is numerically challenging on the length scales we are interested in, and currently beyond computational reach.

The second approach is a modified MHD treatment that neglects the inertia of the ions and treats ambipolar diffusion as an extra diffusive term in the magnetic induction equation \citep{mac1995incorporation}. It assumes that ions stream relative to the neutrals at an instantaneous velocity given by
\begin{equation} \label{eq:driftvel}
    \driftvel = \frac{(\nabla \times {\bm B})\times{\bm B}}{4\pi \gamma_{\rm AD} \rho_n \rho_i},
\end{equation}
where  ${\bm B}$ is the magnetic field, $\gamma_{\rm AD} = \langle \sigma v \rangle/(m_i + m_n )$ is the ion-neutral coupling constant, $\rho_n$ is the density of neutral species, and $\rho_i$ is the density of ions.
Here $\langle\sigma v\rangle$ is the momentum transfer rate coefficient, and $m_i$ and $m_n$ are the ion and neutral mass per particle. The magnetic field would be evolved according to
\beq
\frac{\partial {\bm B}}{\partial t}
=
\nabla \times 
\left( ( {\bm v}_n+{\bm v}_d ) \times {\bm B}\right)
~~~.
\eeq

This second approach is prohibitively expensive for this problem as well. For the 20 pc scales we are interested in, the ambipolar diffusion length is very small in the highest density regions. As a result, to resolve the effects of ambipolar diffusion in these regions would require extremely high resolution. Further, the time-step required for numerical stability in this method scales as $\Delta t_{\rm AD} \propto (\Delta x)^2$, the spatial resolution squared. The combination of these factors make this approach infeasible for our problem.

The third approach is to assume that ideal MHD can be used to evolve the density, fluid velocity, and magnetic field. An additional approximation often made is that the ion-neutral drift velocity $v_d$ can also be approximated by Eq.~\ref{eq:driftvel} with ${\bm B}$ taken to be the field computed assuming ideal MHD. This approach relies on estimating the effects of ambipolar diffusion in post-processing, rather than self-consistently in real time. 
Our approach is most similar to this third approach, with some important modifications. 

\begin{figure}
\begin{center}
	\includegraphics[width=1\columnwidth]{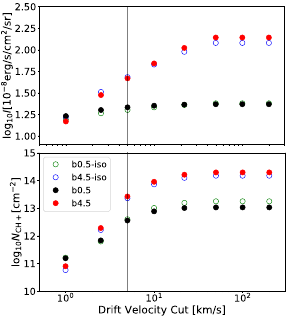}
    \caption{The mean CH$^+$ column density $N_{{\rm CH}^+}$ and total $\HH$ rotational line intensity $I(\HH)$ (defined in sec.~\ref{sec:emission})in each simulation as a function of the cutoff in the ion-neutral drift velocity (see Sec. ~\ref{sec:temp}). The value we adopt in our analysis, 5 km/s, limits us to only counting regions where the ambipolar diffusion Reynolds number is greater than order unity ($v_d \lesssim \sigma_{\rm 3D}$), as well as ensures that the ambipolar diffusion heating is at most of order the driving energy rate. We show this 5 km/s cut as a grayed out vertical line.
    }
    \label{fig:cutoffs}
\end{center}
\end{figure}

Naively assuming that
Eq.~\ref{eq:driftvel} accurately reflects the ion-neutral drift velocities everywhere in a simulation volume leads to several issues.
The
volume-averaged heating rate calculated with Eq.~\ref{eq:driftvel} can easily exceed the volume-averaged driving power, which is unphysical. The heating power per mass may be very large in low density regions, where the drift velocities given by Eq.~\ref{eq:driftvel} can become very large
(see Figure \ref{fig:driftvels}).
A conservative way of dealing with these high drift velocities and heating rates is to exclude 
from our analysis
regions where the drift velocity exceeds some 
chosen
threshold. The exact value of this cut will have an effect on our results. Figure~\ref{fig:cutoffs} examines the effect that this cut has on our results.

%
%%%%%%%%%%%%%%%%%%%%%%%%%%%%%%%%%%%%%%%%%%%%%%%%%
%%%%%%%%%%%%%%%%%%%%%%%%%%%%%%%%%%%%%%%%%%%%%%%%%
%

\section{Heating processes}\label{sec:heat}
As we have simulations with both an isothermal equation of state and a non-isothermal equation of state, we must incorporate heating in two separate ways. 

The first (used for the isothermal equation of state) is to determine the temperature in each cell after the simulation is completed assuming a balance between all cooling and heating processes. Formally, a temperature $T$ is computed in each cell where
\begin{equation}
    \Lambda(\rho,T) = \Gamma(\rho,{\bm v}),
\end{equation}
where $\Lambda(\rho,T)$ (described in Sec.~\ref{sec:cool}) is the total cooling in that cell and $\Gamma(\rho, {\bm v})$ is the total heating including cosmic ray and photoelectric heating and an estimate for the viscous heating (see Secs.~\ref{sec:crpe} and~\ref{sec:vischeat}). Ambipolar diffusion heating is treated differently, and described in Sec.~\ref{sec:adheat}.

The second is to use the explicit integration scheme implemented in the MHD code Athena++ to include the extra heating and cooling terms (described in Secs.~\ref{sec:crpe},~\ref{sec:cool}) dynamically throughout the time domain of the simulation. 
This method
implicitly handles both viscous and shock heating, and has the advantage of allowing for the possibility of adiabatic heating and cooling. However, this can also make the method prohibitively expensive. Occasionally, rarefactions in already cold (dense) gas will adiabatically cool gas well below the equilibrium temperature (which may be as cold as $\approx$10 K in the case of gas with $n_{\rm H} \sim 10^4{\rm cm}^{-3}$). For heating that is proportional to density (as in cosmic ray and photoelectric heating, described below in Sec.~\ref{sec:crpe},~\ref{sec:vischeat}), 
such dense gas may have 
a large heating rate, but a small internal energy, leading to 
a restrictive thermal time step. In this case, we impose a limit on how cold the gas can get, and suppose that it cannot cool below 2.7 K. While this means we effectively inject energy into these few extremely cold regions, the amount is small compared to the other heating terms, and so should have a negligible impact on the global dynamics and chemistry of the simulation.

\subsection{Cosmic ray \& photoelectric heating}
\label{sec:crpe}
Cosmic ray ionizations and the photoelectric effect on dust grains both serve to heat the cold neutral medium.
%in functionally similar ways. 

The heating due to cosmic ray ionizations is a combination of the cosmic ray ionization rate per hydrogen nucleus $\zeta_{\rm H}$, the heat per ionization $\Delta Q$, and the density $n_{\rm H}$.
We take $\zeta_{\rm H} = 1.8\times 10^{-16} {\rm s}^{-1}$ from \citet{indriolo2012investigating}
and $\Delta Q = 10$ eV from \citet{glassgold2012cosmic}.

Our chosen $\Delta Q$ and $\zeta_{\rm H}$
are identical to those chosen in MML15 so that our results are as directly comparable as possible. 
The cosmic ray heating is thus
\begin{equation}
    \Gamma_{\rm CR} = \zeta_{\rm H} \Delta Q n_{\rm H} = 1.9\times 10^{-25}\bigg(\frac{n_{\rm H}}{\langle n_{\rm H}\rangle}\bigg)\,{\rm erg}\,{\rm cm}^{-3}\,{\rm s}^{-1}.
\end{equation}
The dust photoelectric heating rate can be written \citep{Wolfire2003neutral}
\begin{equation}
    \Gamma_{\rm PE} = 1.3\times10^{-24}n_{\rm H}\epsilon G_0 \,{\rm ergs}\,{\rm cm}^{-3}\,{\rm s}^{-1},
\end{equation}
 with $\epsilon$ being the heating efficiency and $G_0$ the FUV intensity in the units of \citet{Habing_1968}. For $G_0 = 1.1$ \citep{Mathis+Mezger+Panagia_1983}, and typical parameters ($\langle n_{\rm H} \rangle = 30\,{\rm cm}^{-3},\, T= 100\,\K,\, x(e) = 1.6\times10^{-4}  $), $\epsilon = 0.018$, and thus
\begin{equation}
    \Gamma_{\rm PE} = 7.6\times 10^{-25}\bigg(\frac{n_{\rm H}}{\langle n_{\rm H} \rangle}\bigg)\,{\rm ergs}\,{\rm cm}^{-3}\,{\rm s}^{-1},
\end{equation}
four times larger than $\Gamma_{\rm CR}$.

\subsection{Viscous heating}\label{sec:vischeat}
While viscous heating is handled implicitly in Athena++ when we use a non-isothermal equation of state, for simulations with an isothermal equation of state we 
 determine the temperature using a post-processing scheme similar to that in \citet{pan2009temperature} and MML15. To estimate the viscous heating in each cell of the simulation, we first compute the rate of shear tensor ${\bm \sigma}$:
\begin{equation}\label{eq:shear}
    {\bm \sigma} = \frac{1}{2}(\nabla{\bm v} + \nabla{\bm v}^T) - \frac{1}{3}(\nabla \cdot {\bm v}){\bm \delta},
\end{equation}
where $T$ denotes the transpose, and ${\bm \delta}$ is the euclidean metric tensor. The gradients here are determined using a cell-centered finite difference method, and so should be second order accurate in the spatial resolution. Given ${\bm \sigma}$, the viscous heating in a cell is 
\begin{equation}\label{eq:vischeat}
    \Gamma_{\nu} = 2\nu \rho {\bm \sigma}:{\bm \sigma}.
\end{equation}
The kinematic viscosity $\nu$ here is not of physical origin; it's a numerical viscosity. Similarly to MML15, we estimate its value by allowing $\nu$ to be a normalization such that the total viscous heating in the simulation is equal to the input driving power; in other words, we assume that all the input driving power goes into viscous heating.

\subsection{Ambipolar diffusion heating}\label{sec:adheat}
We do not include ambipolar diffusion in the dynamics of our simulation. However, in low density regions, ambipolar diffusion can become an important heating mechanism. In order to best estimate what the thermal effects of ambipolar diffusion \textit{would be} in our simulations, we begin with the temperature $T$ either from the code directly (as is the case with simulations b0.5 and b4.5), or post-processed following a procedure similar to that given in MML15 (done for the isothermal simulations b0.5-iso and b4.5-iso). Given these temperatures $T$, we compute a new temperature $\TAD$ by solving the following equation in each cell of the simulation:
 \begin{equation}
     \Lambda(\rho,\TAD) =\Gamma_{\rm AD}(\rho,{\bm B}) + \Lambda(\rho,T),\label{eq:tad}
 \end{equation}
 where $\Gamma_{\rm AD}$ is the heating rate per volume due to ion-neutral friction. This may be expressed as
 \begin{equation}
     \Gamma_{{\rm AD}} = \gamma_{{\rm AD}}\rho_{n} \rho_{i} v_{d}^{2},
 \end{equation}
 where $\gamma_{\rm AD} = \langle \sigma v\rangle/(m_n + m_i)$ is the ion-neutral coupling coefficient (taken to be between C$^+$ ions and $\HH$, identical to that in Eq.~\ref{eq:driftvel}), and $v_d$ is the ion-neutral drift velocity (see Eq.~\ref{eq:driftvel}). 
This assumes instantaneous balance between heating and cooling, which will only be approximately true. It is important to note, however, that this does not throw out our hard-earned dynamical heating, as the new temperature $\TAD$ is bounded by the old temperature $T$ from below, or
 \begin{equation}
     \TAD \ge T.
 \end{equation}

%%%%%%%%%%%%%%%%%%%%%%%%%%%%%%%%%%%%%%%%%%%%%%%%%%%%%%%%%%%%%%%%%%%%%%%
\begin{figure}
\begin{center}
\includegraphics[width=\columnwidth,angle=0,
                 clip=true,trim=0.5cm 4.5cm 0.5cm 0.5cm]%l b r t
{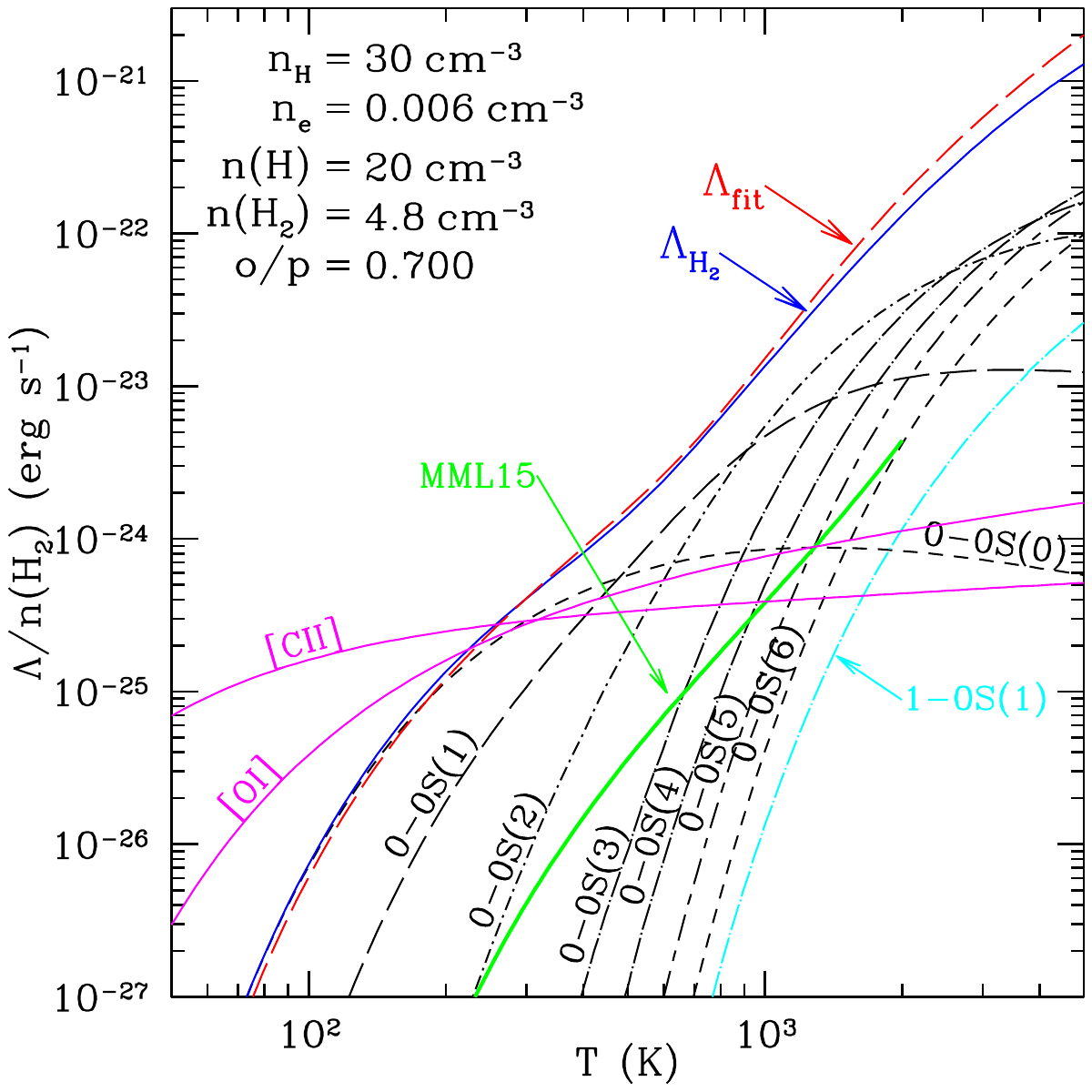}
\caption{\label{fig:MML15} 
%\footnotesize
         Solid line: $\Lambda_{\HH}/n(\HH)$ power radiated per $\HH$
         for statistical equilibrium
         (Eq.\ \ref{eq:pseudoss}),
         Red curve: $\Lambda_{\rm fit}/n(\HH)$ (Eq.\ \ref{eq:fit}).
         Contributions to the cooling of selected lines are shown.
         The magenta curves show the radiated power (per $\HH$)
         from [\ion{C}{II}]158$\micron$ and [\ion{O}{I}]63$\micron$.
         Also shown (green curve)
         is the $\HH$ cooling per $\HH$ given by
         MML15.  Our cooling function $\Lambda_{\HH}$ exceeds
         the MML15 $\HH$ cooling function by a factor $\sim40$
         for $T\gtsim 500\K$.
         }
\end{center}
\end{figure}
%%%%%%%%%%%%%%%%%%%%%%%%%%%%%%%%%%%%%%%%%%%%%%%%%%%%%%%%%%%%%%%%%%%%%%%%

\section{Cooling processes}\label{sec:cool}
\subsection{Fine Structure Lines}\label{sec:coolfine}
In the cold neutral medium, the main avenues for cooling we expect are due to the species $\Cp$, O, and $\HH$. For $\Cp$ and O, we use
\begin{align}
    \Cpcool/n_{\rm H}^{2} &= 3.6 \times 10^{-27} e^{-92\text{ K}/T} \text{erg cm}^{3} \text{ s}^{-1} \label{eq:cpcool} \\
    \Ocool/n_{\rm H}^{2} &= 2.35 \times 10^{-27}\bigg(\frac{T}{100 {\rm K}}\bigg)^{0.4} e^{-228\text{ K}/T} \text{erg cm}^{3} \text{ s}^{-1}. \label{eq:ocool}
\end{align}
These values have been scaled from their original values in \citet{Wolfire2003neutral} to reflect our C and O abundances. 
%For H$_2$, we use the fitting function presented in Eq.\ (\ref{eq:H2cool}).
%(see Sec.~\ref{sec:approx}). 
%As our simulations evolve, these source terms are integrated explicitly.
\subsection{$\HH$ Line Emission}\label{sec:lineemission}

%%%%%%%%%%%%%%%%%%%%%%%%%%%%%%%%%%%%%%%%%%%%%%%%%%%%%%%%%%%%%%%%%%%%%%%
\begin{figure*}
\begin{center}
\includegraphics[width=8.0cm,angle=0,
                 clip=true,trim=0.5cm 4.5cm 0.5cm 0.5cm]%l b r t
                {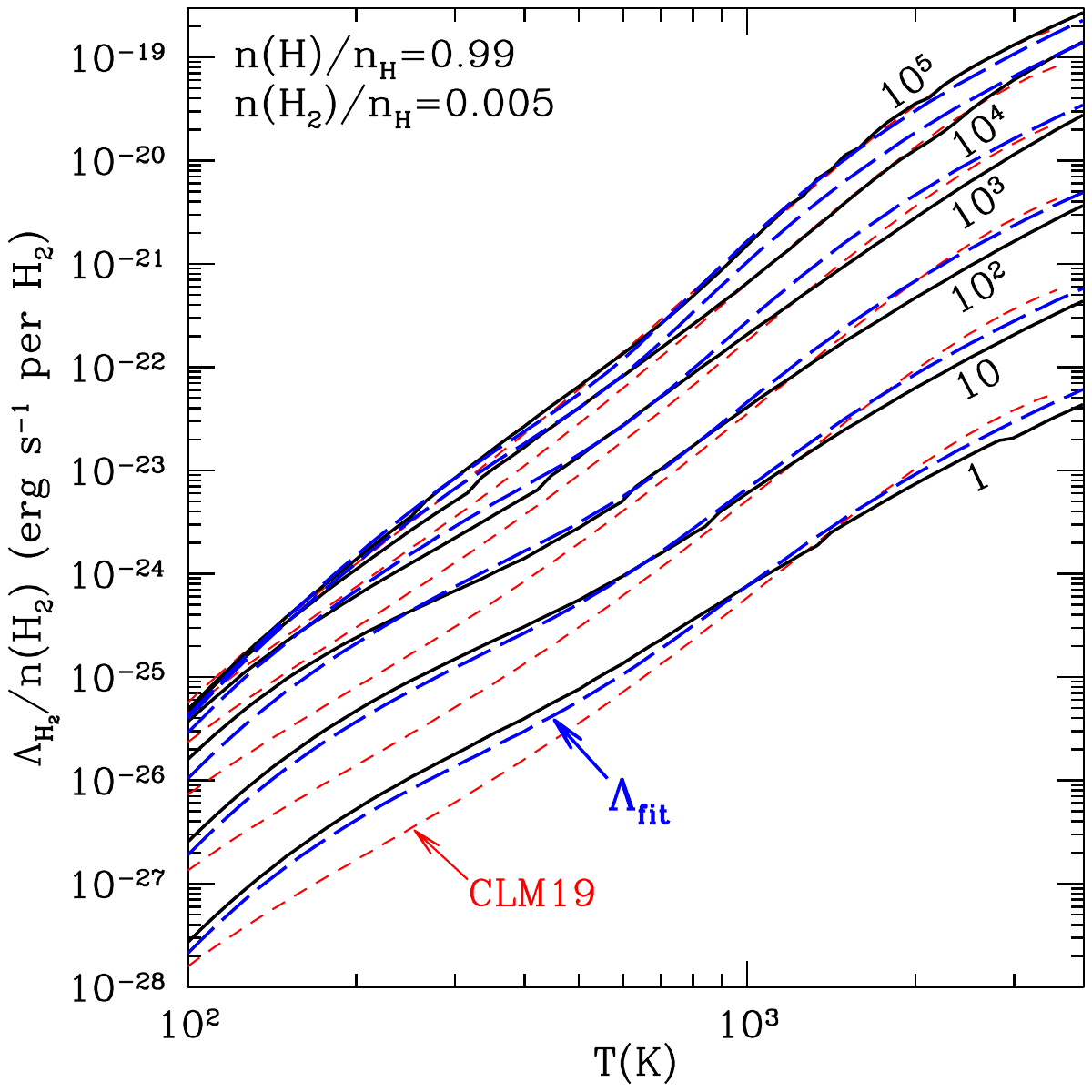}
\hspace*{-0.2cm}\includegraphics[width=8.0cm,angle=0,
                 clip=true,trim=0.5cm 4.5cm 0.5cm 0.5cm]%l b r t
                {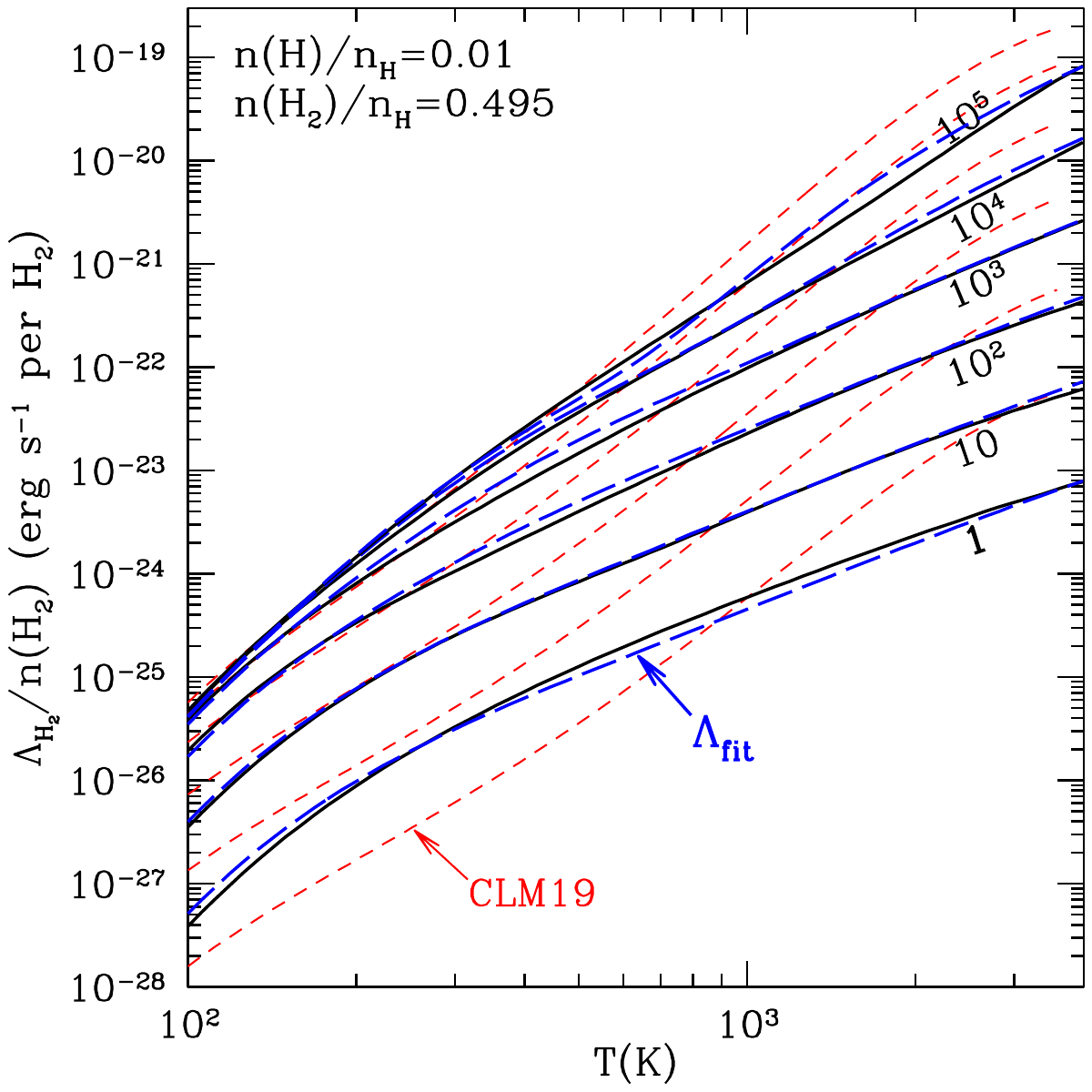}
\caption{\label{fig:Lambda}
%\footnotesize
$\HH$ cooling power
per $\HH$ for densities $\nH=1$, $10$, $10^2$, $10^3$, $10^4$, and
$10^5\cm^{-3}$.
(a) Gas which is 99\% atomic and  1\% $\HH$
(b) Gas which is 99\% $\HH$ and 1\% atomic.
Solid curves: present calculations.
Blue dashed curves: fitting function $\Lambda_{\rm fit}$ (Eq.\ \ref{eq:fit}). Red dashed curves: CLM19 fitting function.
$\Lambda_{\rm fit}$ (Eq.\ \ref{eq:fit}) is much closer to our exact calculation than the CLM19 fit.
Differences are most pronounced for gas that is mainly $\HH$.
}
\end{center}
\end{figure*}
%%%%%%%%%%%%%%%%%%%%%%%%%%%%%%%%%%%%%%%%%%%%%%%%%%%%%%%%%%%%%%%%%%%%%%%

The rotation-vibration lines of $\HH$ can be important cooling channels.
We employ
a new $\HH$ cooling function that is easy to evaluate, but which provides a good approximation to detailed calculations of $\HH$ excitation over a wide range of densities, temperatures, and molecular fractions. 
Our new cooling function $\Lambda_{\HH}$ is based on up-to-date collisional rate coefficients, as described in Appendix \ref{app:hh}.
Figure \ref{fig:MML15} shows the cooling rate for $\nH=30\cm^{-3}$
and molecular fraction $2n(\HH)/\nH=0.3$ that MML15
took to be a representative example for
gas in a turbulent molecular cloud.
We fix the ortho/para ratio at the value $0.7$
adopted by
MML15.  The $\HH$ cooling for this case is shown in Figure \ref{fig:MML15}.

The total $\HH$ cooling is shown (blue solid curve) for
$100\K\leq T\leq 5000\K$.
Also shown are the powers in selected emission lines.
For $T\leq 10^3\K$ the $\HH$ cooling is dominated by 4 rotational lines:
$0-0S(0) 28.22\micron$, $0-0S(1) 17.03\micron$, $0-0S(2) 12.28\micron$,
and $0-0S(3) 9.66\micron$.
At temperatures $T\gtsim 1000\K$,
rotational lines from $J>5$ become important,
and for $T> 2000\K$ the vibrational transitions (e.g.,
$1-0S(1)2.122\micron$) begin to make a significant
contribution to the total cooling.

In Figure \ref{fig:MML15} we also show the cooling power (per $\HH$ molecule) 
in the [CII]158$\micron$ and [OI]63$\micron$ fine structure lines.
We see that for the conditions considered in Figure \ref{fig:MML15},
the fine structure lines dominate the cooling for $T\ltsim 300\K$, but for
$T\gtsim 300\K$ the cooling is dominated by $\HH$.

Figure \ref{fig:MML15} also shows the $\HH$ cooling function from MML15, for
$T\ltsim 2000\K$.
The MML15 cooling function is smaller than the present 
$\Lambda_{\HH}$ by a factor $\sim40$. 

The resulting cooling power per $\HH$, $\Lambda/n(\HH)$, is shown in Figure
\ref{fig:Lambda} for selected densities $1 \leq \nH \leq 10^5\cm^{-3}$, for gas that is predominantly atomic (Fig.\ \ref{fig:Lambda}a) and predominantly $\HH$
(Fig.\ \ref{fig:Lambda}b).  
The atomic and molecular cases differ because the 
rate coefficients for collisional excitation of $\HH$ by $\Ha$ and by $\HH$
can in some cases
differ by large factors.  For example, at $T=5000\K$ the cooling power
per $\HH$ is $\sim5$ times larger in atomic than in molecular gas with the
same $\nH$.

At densities $\nH\ltsim 10^2\cm^{-3}$, $\Lambda_{\HH}/n(\HH)$ 
is approximately linear in $\nH$, with almost all collisional excitations
followed by radiative decay.  At high densities $\nH\gtsim 10^5\cm^{-3}$ the
level populations approach LTE, and $\Lambda_{\HH}/n(\HH)$ becomes
independent of density.

For computational purposes, it is useful to have an 
analytic function $\Lambda_{\rm fit}(n,T)$
that provides an acceptable approximation to the 
``exact'' $\HH$ cooling function 
$\Lambda_{\HH}(T)$ \newtext{for $T\ltsim 5000$K}:
\beqa  \label{eq:fit}
\Lambda_{\rm fit}\hspace{-0.25cm} &=& \hspace{-0.25cm}n(\HH) \sum_{i=1}^4 f_i(n,T) 
\label{eq:H2cool}\\ \nonumber
f_1(n,T) \hspace{-0.25cm}&=&\hspace{-0.25cm} 1.1\xtimes10^{-25} T_3^{0.5} e^{-0.51/T_3}
\left[\frac{0.7 x_1}{1+x_1/n_1} + \frac{0.3 x_1}{1 + x_1/(10n_1)}\right]
\erg\s^{-1}
\\ \nonumber
f_2(n,T)\hspace{-0.25cm} &=& \hspace{-0.25cm}2.0\xtimes10^{-25} T_3 e^{-1/T_3}
\left[\frac{0.35 x_2}{1+x_2/n_2} + \frac{0.65 x_2}{1 + x_2/(10n_2)}\right]
\erg\s^{-1}
\\ \nonumber
f_3(n,T)\hspace{-0.25cm} &=& \hspace{-0.25cm}2.4\xtimes10^{-24} T_3^{1.5} e^{-2/T_3}
\left[\frac{x_3}{1+x_3/n_3}\right]
\erg\s^{-1}
\\ \nonumber
f_4(n,T)\hspace{-0.25cm} &=&\hspace{-0.25cm} 1.7\xtimes10^{-23} T_3^{1.5} e^{-4/T_3}
\left[\frac{0.45 x_4}{1+x_4/n_4} + \frac{0.55 x_4}{1 + x_4/(10n_4)}\right]
\erg\s^{-1}
\\ \label{eq:x1}
x_1 \hspace{-0.25cm}&\equiv&\hspace{-0.25cm} n(\Ha)+5.0 n(\HH)\hspace*{0.5cm} n_1=50\cm^{-3}
\\ \label{eq:x2}
x_2 \hspace{-0.25cm}&\equiv&\hspace{-0.25cm} n(\Ha)+4.5 n(\HH)\hspace*{0.5cm} n_2=450\cm^{-3}
\\ \label{eq:x3}
x_3 \hspace{-0.25cm}&\equiv&\hspace{-0.25cm} n(\Ha)+0.75n(\HH)\hspace*{0.5cm} n_3=25\cm^{-3}
\\ \label{eq:x4}
x_4 \hspace{-0.25cm}&\equiv&\hspace{-0.25cm} n(\Ha)+0.05n(\HH)\hspace*{0.5cm} n_4=900\cm^{-3}
\eeqa
The coefficients multiplying $n(\HH)$ in Eq.(\ref{eq:x1}-\ref{eq:x4})
reflect the collisional rate coefficients for excitation by
$\HH$ relative to excitation by H and He.
We see that Eq.\ (\ref{eq:fit}) provides a fairly good fit to $\Lambda$ over
a wide range of temperatures and densities, for both atomic gas
(Fig.\ \ref{fig:Lambda}a) and molecular gas
(Fig.\ \ref{fig:Lambda}b).

\citet{Coppola+Lique+Mazzia+etal_2019} (hereafter CLM19) provide a
fitting function for $\HH$ cooling over the $10^2\K$ -- $4000\K$
temperature range.
Figure \ref{fig:Lambda} compares the CLM19 fitting
function to our calculated $\HH$ cooling rates in predominantly atomic gas: the CLM19 fitting function tends to underestimate our computed cooling rates by factors of $\sim 2$ for $T\lesssim 500\K$.

Both CLM19 and the present study use H-H$_2$ collision cross sections from \citet{Lique_2015}. The difference in the $T\ltsim 500\K$ cooling appears to be
due to differences in adopted rates for collisional excitation by He:
\citet{Coppola+Lique+Mazzia+etal_2019} used quasi-classical trajectory
cross sections from \citet{Celiberto+Capitelli+Colonna+etal_2017}
whereas we use quantum-mechanical results from
\citet{LeBourlot+PineaudesForets+Flower_1999}, which are believed to be
more accurate at low energies.

For molecular gas, the CLM19 fitting function underestimates the cooling rate by a factor of $\sim 3$ for temperatures $\lesssim 500\K$, and overestimates it by a similar factor for $\gtrsim 500\K$. Both the present study and CLM19 assume ortho-para equilibration,
resulting in a low ortho-para ratio at the lower temperatures.

\section{CH$^+$ Chemistry}\label{sec:chplus}

\begin{figure}
\begin{center}
	\includegraphics[width=1\columnwidth]{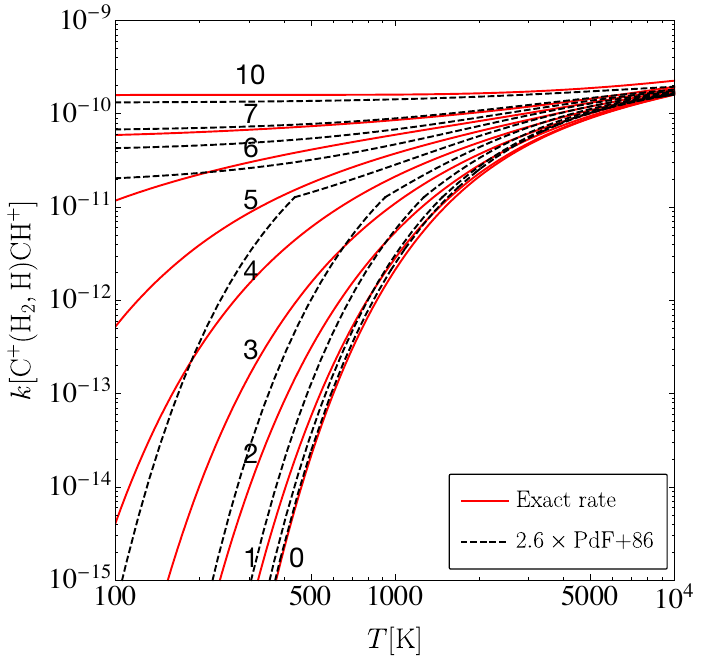}
    \caption{The exact rate coefficient (red, solid) for the reaction~(\ref{eq:react}) (see text)
    and the approximation (black, dashed) from \citet{Pineau_des_Forets+Flower+Hartquist+Dalgarno_1986}, but multiplied by 2.6 to better match the exact rate. The exact curves are labeled by the ion-neutral drift velocity $v_d$ (km/s). The approximate curves correspond sequentially to these same drift velocities. We see that the approximation is more accurate at higher values of the rate coefficient, with the largest discrepancies being when the drift velocities are 5 and 6 km/s. }
    \label{fig:ratecoeff}
\end{center}
\end{figure}

 When ions are streaming through the neutrals, the rate for the endothermic reaction (\ref{eq:react}) is affected by the nonthermal distribution of ion-neutral impact speeds.  Following \citet{Flower+Pineau_des_Foret+Hartquist_1985} we employ a rate coefficient which is a function of an effective temperature
\begin{equation}
    T_{\rm eff} = \TAD + \frac{\mu}{3k}v_d^2
    =\TAD + 10^3{\rm K}\left(\frac{v_d}{3.8\,{\rm km\,s}^{-1}}\right)^2\label{eq:teff},
\end{equation}
where $\mu$ is the reduced mass of C$^+$ and $\HH$ and $k$ is the Boltzmann constant. This applies when $T_{\rm eff}\ge 1547\K$ \citep{Pineau_des_Forets+Flower+Hartquist+Dalgarno_1986}. When $T_{\rm eff} < 1547\K$, we instead have that the reaction rate is a function of
\begin{equation}
    T' \equiv \TAD \frac{4640\K}{4640\K-\mu v_d^2/k}\approx \frac{\TAD }{1-(v_d/4.7\,{\rm km/s})^2}
\end{equation}
As with MML15, we assume instantaneous balance between formation and destruction of CH$^+$. The rate coefficients for each of the various creation and destruction mechanisms are
 \begin{align}
     {\rm C}^+ + \HH &\rightarrow {\rm CH}^+ + {\rm H}; &k_{{\rm CH}^+} &= 2.6\times 10^{-10} \exp[-\xi]\,{\rm cm}^3\,{\rm s}^{-1},\nonumber\\
     &&\xi&\equiv \text{max}\bigg\{\frac{4640\K}{\Teff},\frac{4640\K}{T'}\bigg\},\nonumber\\
     {\rm C}^+ + {\rm H} &\rightarrow {\rm CH}^+; &k_{{\rm ra}} &= 4.46\times 10^{-17} T_2^{-1/2}\times\nonumber\\
     &&&\exp[-0.229 T_2^{-2/3}]\,{\rm cm}^3\,{\rm s}^{-1},\nonumber\\
      {\rm CH}^+ + {\rm H}&\rightarrow {\rm C}^+ + \HH;  &k_{{\rm HI}} &= 1.5\times 10^{-10}\,{\rm cm}^3\,{\rm s}^{-1}, \nonumber\\
      {\rm CH}^+ + \HH &\rightarrow {\rm CH}_2^+ + {\rm H};  &k_{\HH} &= 1.2\times 10^{-9}\,{\rm cm}^3\,{\rm s}^{-1}, \nonumber\\
      {\rm CH}^+ + e &\rightarrow {\rm C} + {\rm H};  &k_{e} &= 5.2\times10^{-8} T_2^{-0.17}\,{\rm cm}^3\,{\rm s}^{-1} %1.5\times 10^{-7}\,{\rm cm}^3\,{\rm s}^{-1}.
 \end{align}
 $k_{{\rm CH}^+}$ is an approximate form given by \citet{Pineau_des_Forets+Flower+Hartquist+Dalgarno_1986}, but increased by a factor 2.6 to better match the exact rate shown in Fig.~\ref{fig:ratecoeff}. 
 In Fig.~\ref{fig:ratecoeff}, we compare the \citet{Pineau_des_Forets+Flower+Hartquist+Dalgarno_1986} approximation against 
 The "exact" rate in Fig.~\ref{fig:ratecoeff} is calculated following \citet{Draine+Katz_1986a} for various $v_d$ using the cross section for the $v=0$, $J=0$ state of $\HH$ 
 \citep{Gerlich+Disch+Scherbarth_1987,2013ApJ...766...80Z}.  $k_{{\rm HI}}$, and $k_{\HH}$ are those from the Meudon PDR code\footnote{https://ism.obspm.fr/?page\_id=33}. $k_{\rm ra}$ is from \citet{Barinovs+vanHemert_2006}, while $k_{e}$ is from \citet{Charkabarti+Mezei+Motapon+etal_2018}. 
 Balancing formation with destruction gives
\begin{equation}
    n_{{\rm CH}^+} = x({\rm C}^+)\frac{x(\HH)k_{{\rm CH}^+}+[1-2x({\HH})]k_{\rm ra}}{[1-2x(\HH)]k_{\rm HI}+ k_{\HH} x(\HH) + k_e x(e)}n_{\rm H}.\label{eq:chp}
\end{equation}
 It is clear from this expression that the CH$^+$ abundance is most sensitive to the abundances of C$^+$ and $\HH$. Solving for the abundance as a function of time when perturbed from equilibrium, one finds that the chemical relaxation time is on the order of 250 years, much shorter than all other time scales in our simulation, so the assumption of equilibrium chemistry is a good one.

%%%%%%%%%%%%%%%%%%%%%%%%%%%%%%%%%%%%%%%%%%%%%%%%%
%%%%%%%%%%%%%%%%%%%%%%%%%%%%%%%%%%%%%%%%%%%%%%%%%
%

\section{Results}\label{sec:results}

\subsection{Temperature \& ion-neutral drift}\label{sec:temp}

%
%%%%%%%%%%%%%%%%%%%%%%%%%%%%%%%%%%%%%%%%%%%%%%%%%
\begin{figure}
\begin{center}
	\includegraphics[width=1\columnwidth]{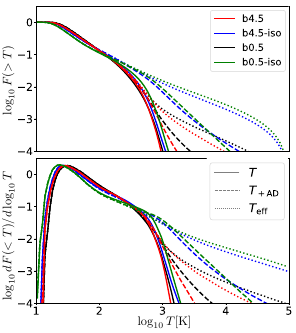}
    \caption{Mass weighted cumulative distribution functions (top) and probability density functions (bottom) of temperature $T$ (solid), temperature adjusted with ambipolar diffusion heating $\TAD$ (dashed, see Eq.~\ref{eq:tad}), and effective temperature (dotted, see Eq.~\ref{eq:teff}) for simulations with two different mean magnetic field strengths and either an isothermal equation of state post-processed as described in Sec.~\ref{sec:heat}. It is clear that the difference between the two procedures is not dramatic. Further, as might be expected, for higher field strengths, the PDFs for $\TAD$ and $T_{\rm eff}$  show a much larger mass of gas at or above 1000K, and thus able to produce CH$^+$ and $\HH$ rotational line emission.}
    \label{fig:temperatures}
\end{center}
\end{figure}
%%%%%%%%%%%%%%%%%%%%%%%%%%%%%%%%%%%%%%%%%%%%%%%%%
%
The chemistry we are primarily interested in takes place in gas with temperatures $\gtrsim 1000$K. Fig.~\ref{fig:temperatures} shows that in our simulations, the
mass-fraction of the gas with either $\TAD\gtrsim 1000\K$ or $\Teff \gtrsim 1000\K$ is $\sim 1\%$, just as \citet{falgarone1995intermittency} and \citet{pan2009temperature} suggested was necessary to produce 
the observed column densities of CH$^+$. If the temperature becomes \textit{too} high, ($\gtrsim 5000$K) molecular hydrogen will be dissociated. We also suppose that CH$^+$ production will cease if the ion-neutral streaming velocity $v_d$ exceeds
$\sim$22 km/s, where the center-of-mass energy is sufficient to dissociate $\HH$ via
${\rm C}^+ + \HH \rightarrow {\rm C}^+ + 2{\rm H}$.

With ideal MHD, we find the drift velocities given by Eq.~\ref{eq:driftvel} can sometimes become very large in low density regions, sometimes reaching values as large as 100\,km/s in very low density regions. In nature, were drift velocities this high produced, they would (1) self-limit by decreasing the magnetic field strength directly through diffusion, and (2) create high ambipolar diffusion heating rates that would drive up the temperature and thus ionization of the gas, leading to stronger ion-neutral coupling and a drop in the drift speed.
We thus suspect that in the cold neutral medium, drift velocities do not routinely reach these high values $v_d \gtrsim 5$ km/s. 
For these reasons, we choose to simply ignore
CH$^+$ production in
regions in our simulation where the drift velocity exceeds 5 km/s. We explore the effect of varying this cutoff velocity in Fig.~\ref{fig:cutoffs}.

%
%%%%%%%%%%%%%%%% f7 %%%%%%%%%%%%%%%%%%%%%%%%%%%%%%%%%
\begin{figure*}
\begin{center}
	\includegraphics[width=0.75\textwidth]{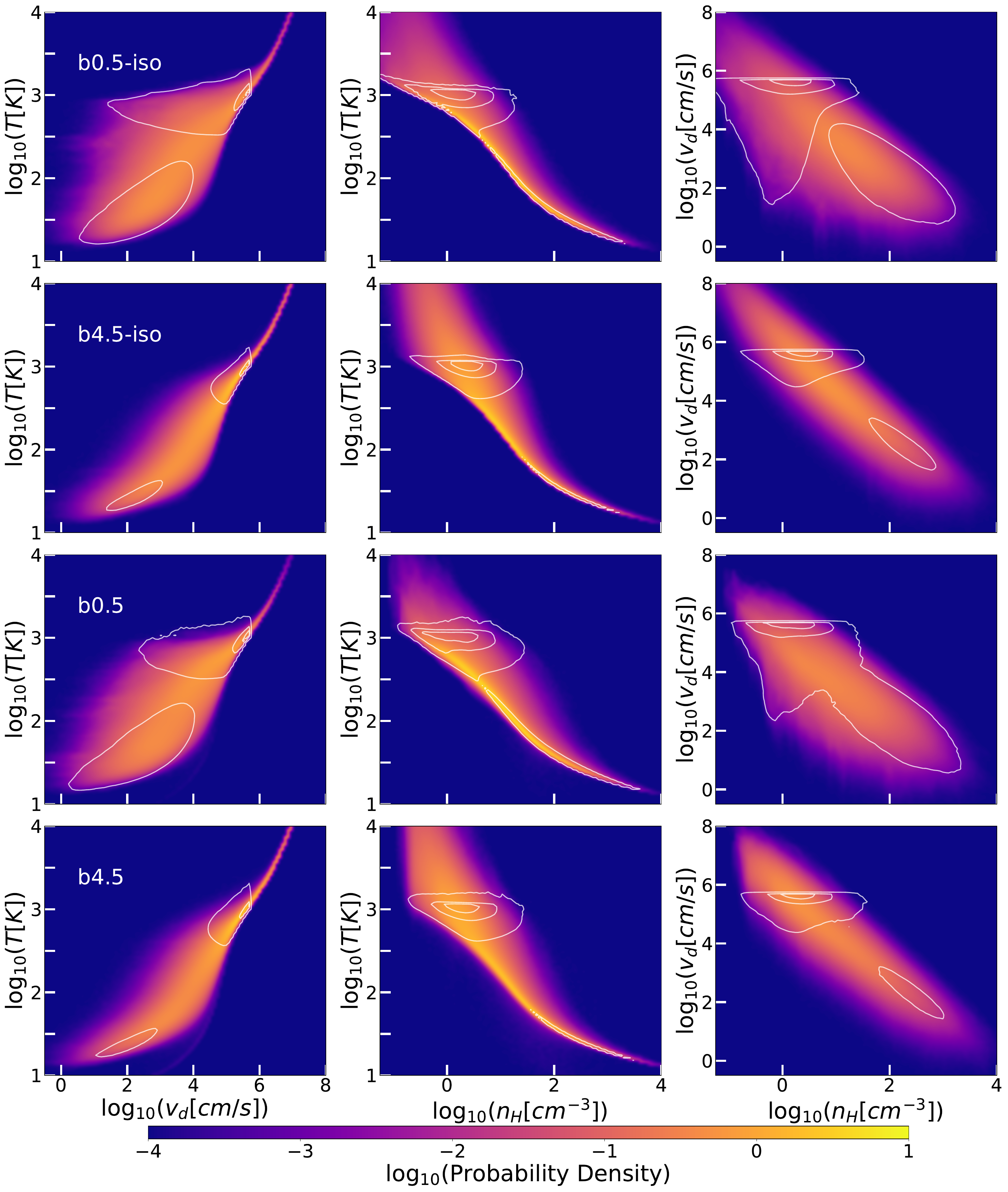}
    \caption{\textbf{Left:} The joint probability density function of  gas temperature T and ion-neutral drift velocity $v_d$ (see Secs.~\ref{sec:AD} and \ref{sec:heat}). The white contours contain 50\%, 90\%, and 99.9\% of the CH$^+$. This is true for the contours across the other columns as well. \textbf{Middle:} Joint PDF for temperature and density $n_{\rm H}$. The scatter above the density dependent equilibrium is due to additional heating due to shear viscosity, shock heating (in the case of b0.5 and b4.5), and ambipolar diffusion. \textbf{Right:} Joint PDF for $v_d$ and $n_{\rm H}$. The temperatures plotted here are those referred to as $\TAD$ in the text; that is, they include ambipolar diffusion heating in their determination.
    }
    \label{fig:512panel}
\end{center}
\end{figure*}
%%%%%%%%%%%%%%%%%%%%%%%%%%%%%%%%%%%%%%%%%%%%%%%%%
%

\subsection{CH$^+$ abundance}\label{sec:chp}

%
%%%%%%%%%%%%%%%%%%%%% f8 %%%%%%%%%%%%%%%%%%%%%%%%%%%%
\begin{figure}
\begin{center}
	\includegraphics[width=1\columnwidth]{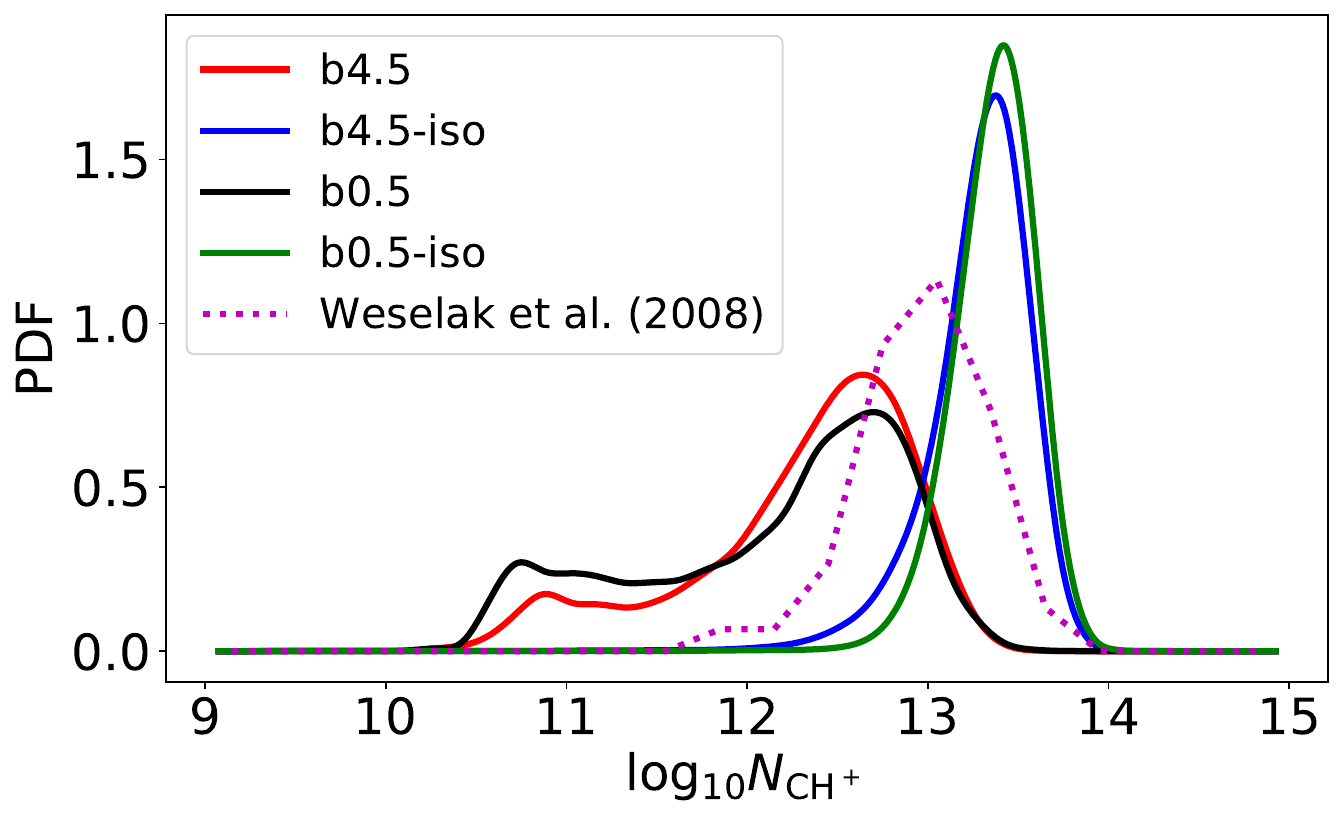}
    \caption{Probability density functions for the (log) column density of CH$^+$ in each of our four simulations, computed as described in Sec.~\ref{sec:chp}. The solid magenta line corresponds to data from \citet{weselak2008relation}. The high magnetic field strength simulations, b4.5 and b4.5-iso somewhat exceed the \citet{weselak2008relation} data. The low field strengths (b0.5 and b0.5-iso) fall short of explaining the observations.
    }
    \label{fig:chp}
\end{center}
\end{figure}
%%%%%%%%%%%%%%%%%%%%%%%%%%%%%%%%%%%%%%%%%%%%%%%%%
%

With the CH$^+$ formation rate depending on the effective temperature $T_{\rm eff}$, we expect the CH$^+$ abundance to be biased towards both high 
$T$ and high ion-neutral drift velocity regions. These regions happen to mostly coincide. In Fig.~\ref{fig:512panel}, we show joint histograms of 
$T$ vs.\ drift velocity, 
$T$ vs.\ density, and drift velocity vs.\ density for each of our four simulations. The white contours in Fig.~\ref{fig:512panel} represent where 99\%, 90\% and 50\% of the CH$^+$ exists on each plot. These contours demonstrate that in all of our simulations, CH$^+$ is produced in low density ($n_{\rm H} \sim 3$ cm$^{-3}$), high drift velocity ($v_d \gtrsim 1$ km/s), hot regions ($T\gtrsim 500$ K).

However, these two modes for creating CH$^+$ are not equal in efficacy. For both the high field strength simulations (b4.5 and b4.5-iso) and the low field strength simulations (b0.5 and b0.5-iso) it seems that high drift velocity regions are responsible for the vast majority of CH$^+$ production. Fig.~\ref{fig:cutoffs} shows that when we exclude regions with $v_d\ge v_{\rm cut}$, as we 
decrease $v_{\rm cut}$, the CH$^+$ abundance in the simulations is attenuated beginning at about $v_{\rm cut} \approx 20$ km/s in all simulations. As we reduce $v_{\rm cut}$ down to 1 km/s 
nearly all of the CH$^+$ vanishes, with our abundances falling to between $10^{10}$ and $10^{11}$ cm$^{-2}$. 

As another way to estimate the effect of ion-neutral drift on CH$^+$, we may use the ordinary temperature $T$ instead of  $\TAD$ or $T_{\rm eff}$ in computing the CH$^+$ abundance using Eq.~\ref{eq:chp}. Doing so, we find median CH$^+$ 
column densities that are lower than those listed in Table~\ref{tab:sims} by a factor of between 16 and 50 (depending on the simulation),
far below those observed in \citet{weselak2008relation}.
This suggests that ion-neutral drift is critical for CH$^+$ formation.

\begin{figure}
\begin{center}
	\includegraphics[width=1\columnwidth]{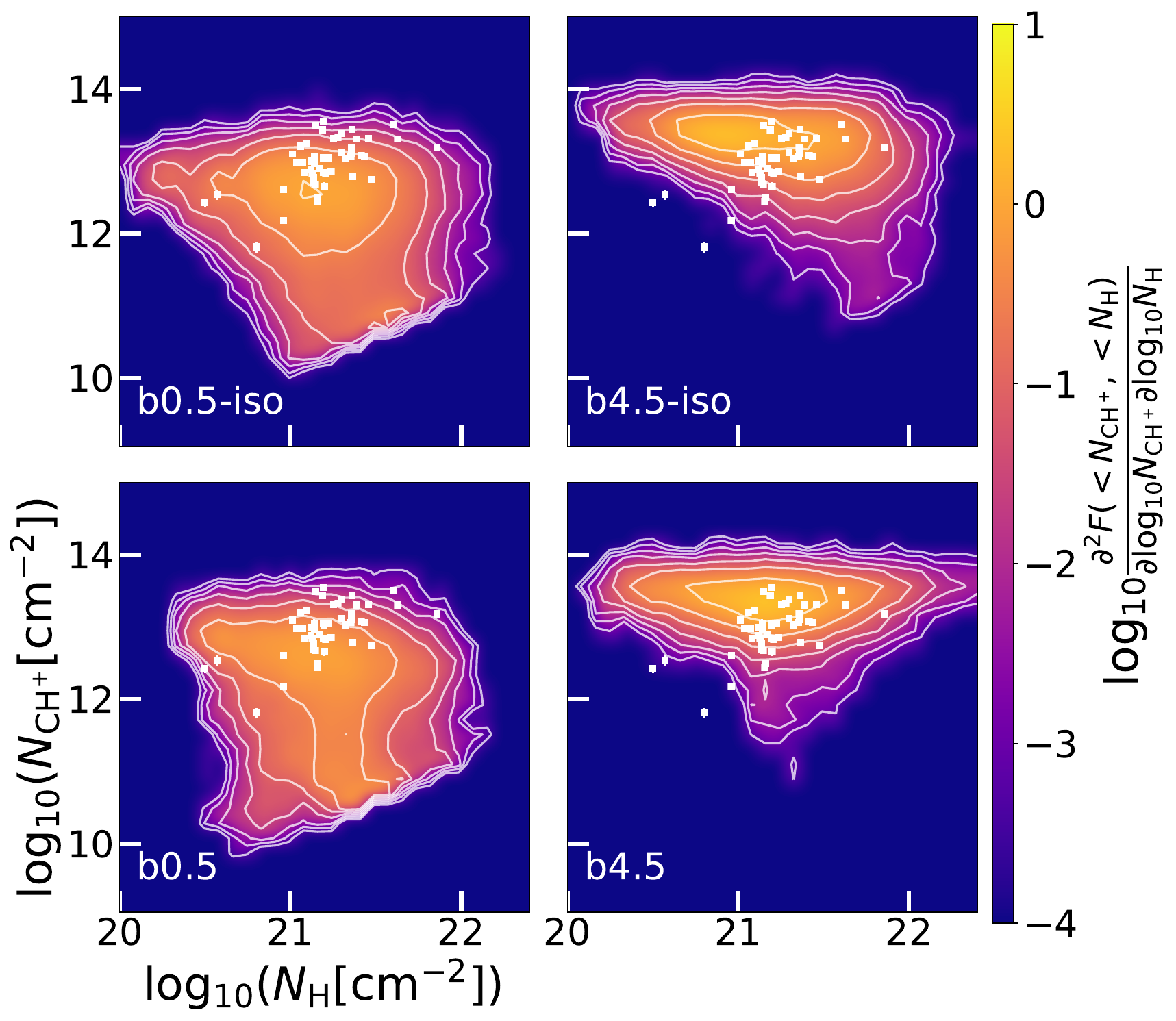}
    \caption{Joint histograms of the column density of CH$^{+}$ and N$_{\rm H}$ for $512^2$ sightlines in each simulation (Note: this plot is insensitive to viewing angle). The white data points are from \citet{weselak2008relation}. Contours reflect the levels of the PDF every 1/2 dex. Our 
    b4.5 model does seem to reproduce the abundances seen in \citet{weselak2008relation}. However, it should be noted that we have 
    omitted CH$^+$ in
    regions where the ion-neutral drift velocity $v_d$ exceeds 5 km/s. 
    }
    \label{fig:chpvsnh}
\end{center}
\end{figure}

In Figs.~\ref{fig:chp} and~\ref{fig:chpvsnh}, we compare the results of our model to \citet{weselak2008relation}'s data. Fig.~\ref{fig:chp} shows reasonable agreement between the \citet{Weselak+Galazutdinov+Musaev+Krelowski_2008} data and our simulations b4.5 and b4.5-iso, though there may be a larger tail to low CH$^+$ column density in the observations than in our simulation data.
In Fig.\ \ref{fig:chpvsnh} we do not find the same correlation between column density of H nuclei and CH$^+$ in our simulations that seems to be present in the \citet{weselak2008relation} sample. 

Our simulations assume a constant $\HH$ fraction, $n(\HH)=0.16n_{\rm H}$ -- we do not follow formation and destruction of $\HH$, which would require treating the radiative transfer of the far-UV photons responsible for photodissociation of $\HH$.  Thus the low $N_{\rm H}$ sightlines in our simulations support CH$^+$ formation via reaction (\ref{eq:react}).
In the real ISM, however, sightlines with
$N(\HH)\ltsim 10^{18}\cm^{-2}$ tend to have low $\HH$ fractions, because photodissociation is insufficiently suppressed by self-shielding
\citep{Draine+Bertoldi_1996} -- this accounts for the absence of CH$^+$ detections for $N_{\rm H}<2.5\times10^{20}\cm^{-2}$ in \citet{weselak2008relation}.

\newtext{Our simulations were limited to a single $20\times20\times20\,{\rm pc}^3$ molecular
region, whereas observations may sample molecular regions that are
both smaller and larger.  However, our simulated $20\times20\times20\,{\rm pc}^3$
simulations are quite clumpy, so that sightlines sample column
densities $N_{\rm H}$ ranging from $10^{20.3}$ to $10^{22}$ cm$^{-2}$ (see
Figs. \ref{fig:column_density}, \ref{fig:chpvsnh}), similar to the range of $N_{\rm H}$ values in the \citet{Weselak+Galazutdinov+Musaev+Krelowski_2008} sample.}
%\newtext{When comparing slightlines that contain observations of both CH$^+$ and H column densities to sightlines in our simulation, there is an implicit assumption made: while our simulations reflect clouds of a single size (20 pc), those similar clouds in nature may vary in size by several orders of magnitude. Thus, when plotting observations over simulation data as in fig.~\ref{fig:chpvsnh}, one must keep in mind the caveat that some of these observations may pass through larger or smaller clouds. Larger clouds may have a higher $\HH/{\rm H}$ fraction due to more efficient $\HH$ self-shielding, which would be more favorable for the formation of CH$^+$. We might expect more scatter in observations of $N({\rm CH}^+)$ at fixed $N_{\rm H}$ than in our simulation data for this reason. }

\begin{figure}
\begin{center}
	\includegraphics[width=1\columnwidth]{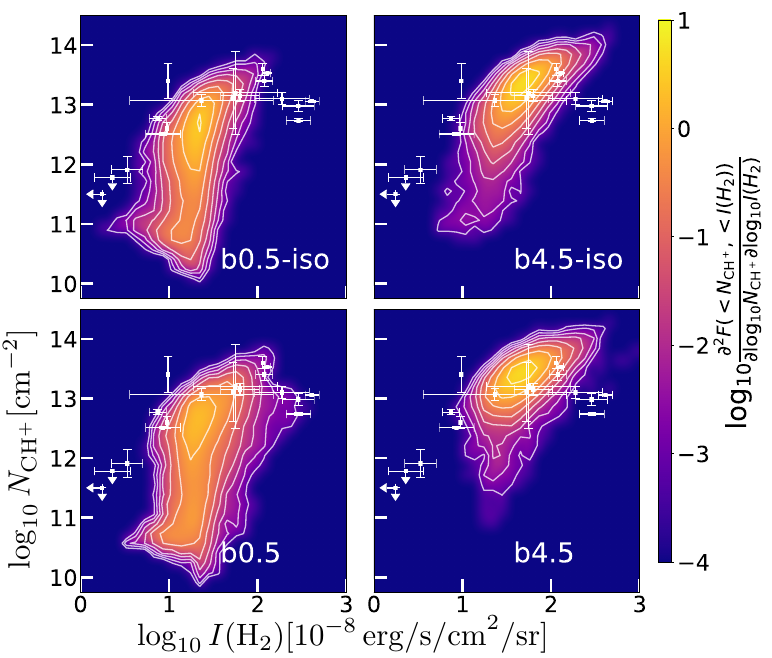}
    \caption{Joint histograms of the column density of CH$^{+}$ and the total intensity in $\HH$ rotational line emission, I$(S(0)+S(1)+S(2))$. A strong correlation exists between the two quantities at higher column densities of CH$^+$. The $\HH$ data are collated from \citet{spitzer1974column}, \citet{snow1976analysis}, \citet{black1977models}, \citet{frisch1980interstellar},\citet{FrischJura1980ApJ}, \citet{Gry+Boulanger+Nehme+etal_2002}, and \citet{Lacour+Ziskin+Hebrard+etal_2005}. The CH$^+$ data are from \citet{hobbs1973interferometric}, \citet{chaffee1975line}, \citet{frisch1979interstellar}, \citet{FrischJura1980ApJ}, \citet{Lambert+Danks_1986}, \citet{Gry+Boulanger+Nehme+etal_2002}, and \citet{weselak2008relation}.
    }
    \label{fig:chpvsem}
\end{center}
\end{figure}

\newtext{In this study, we define $I(\HH)$ as the intensity of emission from the first three $\HH$ rotational lines 0-0 $S(0)$, $S(1)$, and $S(2)$. How it is computed from the simulations and observational data is specified in section~\ref{sec:emission}.}
Comparing $N_{{\rm CH}^+}$ to $I(\HH)$, Fig.~\ref{fig:chpvsem} shows a strong correlation between these quantities, in particular where column densities of CH$^+$ are high or where $I(\HH)$ is high.
This is in line with the 
observed correlation of
rotationally-excited $\HH$ 
 with $N_{{\rm CH}^+}$ \citep{Lambert+Danks_1986,FrischJura1980ApJ,spitzer1974column}.

\begin{figure}
\begin{center}
	\includegraphics[width=1\columnwidth]{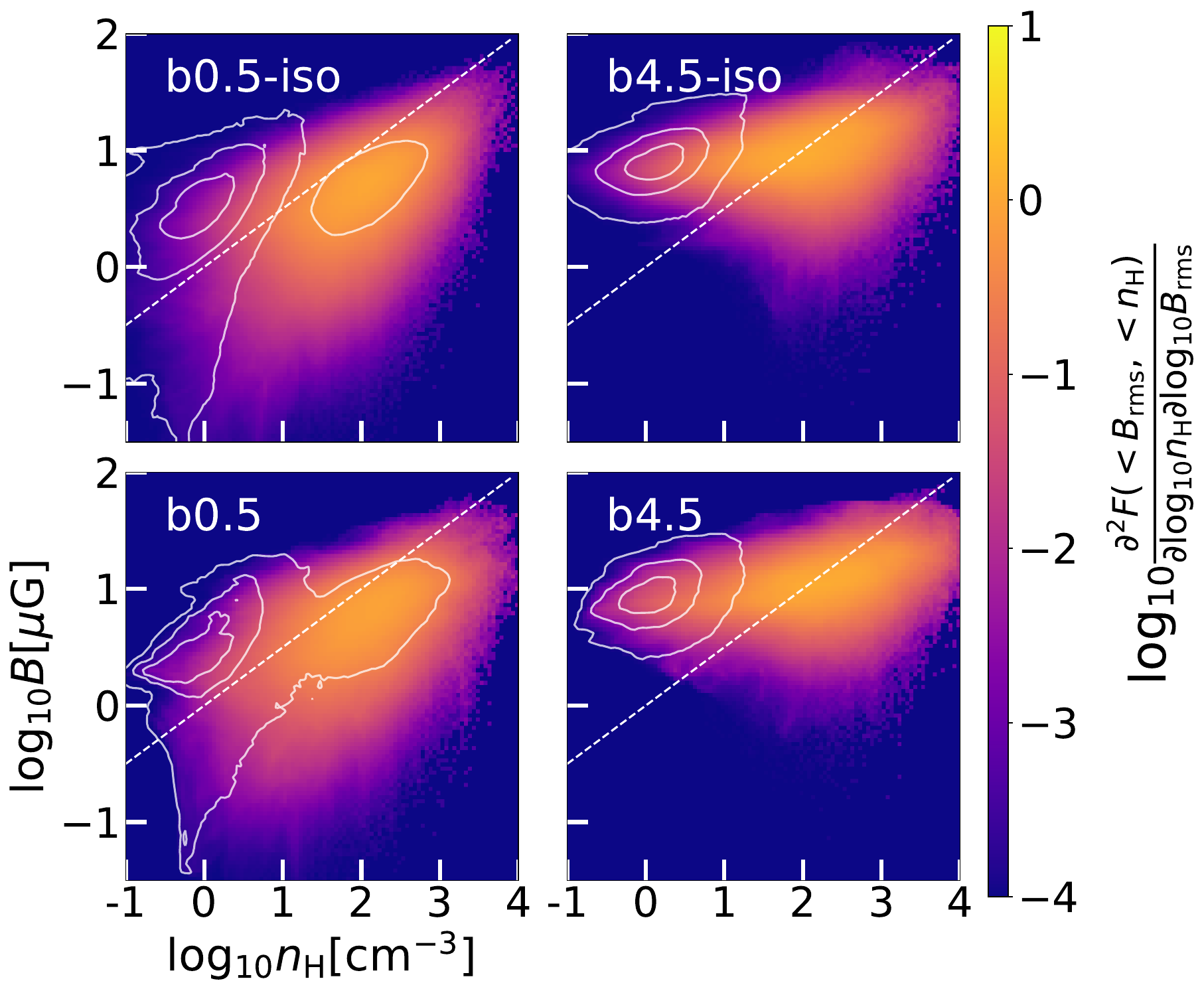}
    \caption{Joint probability densities of the (log) magnetic field at each position in our simulations versus the (log) density of hydrogen nuclei. Contours represent 50\%, 90\% and 99.9\% of the CH$^+$ in the volume. The dashed line represents the scaling $B\propto \sqrt{\rho}$. 
    For $B_0=0.5\muG$ simulations where most of the field strength is generated by the turbulence and $\mathcal{M}_A \approx 3$ (b0.5, b0.5-iso), the scaling $B\propto \sqrt{\rho}$ is approximately followed, albeit with large scatter. However, for $B_0=4.5\muG$ simulations 
    with $\mathcal{M}_A \approx 1$ (b4.5 and b4.5-iso), there is not much dependence of the magnetic field on density. 
    }
    \label{fig:bvsnh}
\end{center}
\end{figure}

Figure~\ref{fig:bvsnh} shows the joint probability density function of magnetic field strength versus number density of hydrogen nuclei, with contours containing 50\%, 90\%, and 99.9\% of the CH$^+$ in our simulation. As CH$^+$ appears to cluster around densities $n_{\rm H} \sim 1\,{\rm cm}^{-3}$ in our simulations, we may consider the CH$^+$ column density to be a probe of the heating processes in the lowest density molecular regions. 
 
%
%%%%%%%%%%%%%%%%% f9 %%%%%%%%%%%%%%%%%%%%%%%%%%%%%%%%
%%%%%%%%%%%%%%%%%%%%%%%%%%%%%%%%%%%%%%%%%%%%%%%%%
%

 \subsection{Synthetic CH$^+$ velocity distributions}\label{sec:chplines}

 We construct synthetic line-of-sight (1D) particle velocity distributions for $\HH$ and CH$^+$ for qualitative comparison with 
 observations \citep[e.g.,][]{pan2004cloud}.
In each cell of a simulation, we have the neutral fluid velocity ${\bm v}$ and the ion-neutral drift velocity ${\bm v}_d$ (calculated as in Eq.~\ref{eq:driftvel}). The fluid velocity of the ions in that cell is
 \begin{equation}
     {\bm v}_i = {\bm v}+{\bm v}_d.
 \end{equation}
We assume Maxwellian velocity distributions,
 centered on ${\bm v}$ for neutrals, and 
 ${\bm v}_i$ for ions. In addition, below the grid scale of the simulation, there will be some microturbulent contribution
 to the velocity dispersions of both the ionic and neutral species. If the power spectrum for each simulation were purely Kolmogorov, this microturbulent contribution would be $\sigma_0 \approx 0.5$ km/s, and this is the value we assume. The total width of the velocity distribution for a species $X$ in a cell $j$ is therefore,
 \begin{equation}
     \sigma^2_{X,j} = \frac{k T_{X,j}}{m_X} + \sigma_0^2, \label{eq:sigma2X}
 \end{equation}
 where $k$ is the Boltzmann constant, $T_{X,j}$ is the kinetic temperature of species $X$ in cell $j$, and $m_X$ is the mass of species $X$.
 
 The velocity profile ${\rm d}N^{(X)}/{\rm d}u$ of a species $X$ at a velocity $u$ is 
 \begin{equation}
     \frac{{\rm d}N^{(X)}}{{\rm d}u} = \Delta x \bigg(\frac{\pi}{2}\bigg)^{1/2}\sum_{j=1}^N \frac{n_j^{(X)}}{\sigma_{X,j}}\exp{\Big(-(u - v_{X,j})^2/2\sigma_{X,j}^2\Big)},
 \end{equation}
 where $n_{X,j}$ is the number density of species $X$ in cell $j$, and $\Delta x$ is the physical size of a cell.

%
%%%%%%%%%%%%%%% f10 %%%%%%%%%%%%%%%%%%%%%%%%%%%%%%%%%%
%%%%%%%%%%%%%%%%%%%%%%%%%%%%%%%%%%%%%%%%%%%%%%%%%
%

%
%%%%%%%%%%% f11 %%%%%%%%%%%%%%%%%%%%%%%%%%%%%%%%%%%%%%
%%%%%%%%%%%%%%%%%%%%%%%%%%%%%%%%%%%%%%%%%%%%%%%%%
%

%
%%%%%%%%%%% f12 %%%%%%%%%%%%%%%%%%%%%%%%%%%%%%%%%%%%%%
\begin{figure*}
\begin{center}
	\includegraphics[width=2\columnwidth]{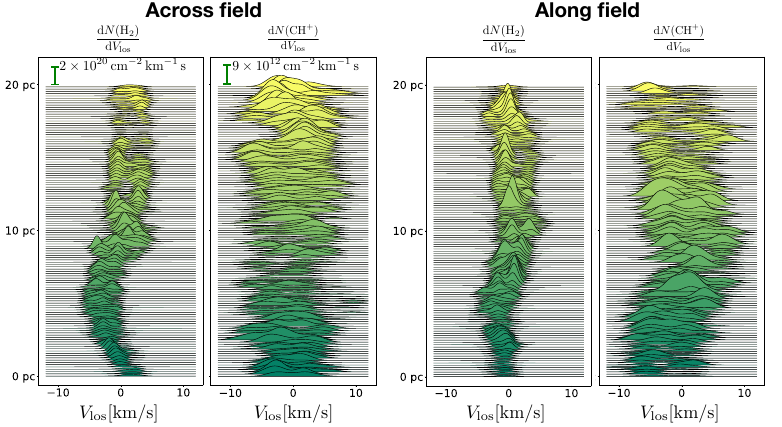}
    \caption{{\bf Left:} Line-of-sight velocity profiles for $\HH$ molecules and CH$^+$ molecules computed looking across the mean magnetic field. These are for the simulation b4.5 (see Tab.~\ref{tab:sims}). The 
    vertical offset increases linearly with a coordinate perpendicular to the line of sight and ${\bf B}_0$. For scale, two green bars are provided to indicate the height of the profiles. In all panels, color is simply shows progression from one side of the simulation to the other. {\bf Right:} The same, but computed looking along ${\bf B}_0$. The scales are the same as the corresponding scale bars on the left. We see considerable diversity among the CH$^+$ line profiles, and no obvious correlation with the $\HH$ profiles. The CH$^+$ profiles are also consistently wider, with a width $\sim 10$ km/s, while the $\HH$ only has a width of at most a few km/s.
    }
    \label{fig:chplines}
\end{center}
\end{figure*}
%%%%%%%%%%%%%%%%%%%%%%%%%%%%%%%%%%%%%%%%%%%%%%%%%
%

%
%%%%%%%%%%% f13 %%%%%%%%%%%%%%%%%%%%%%%%%%%%%%%%%%%%%%
\begin{figure}
\begin{center}
	\includegraphics[width=1\columnwidth]{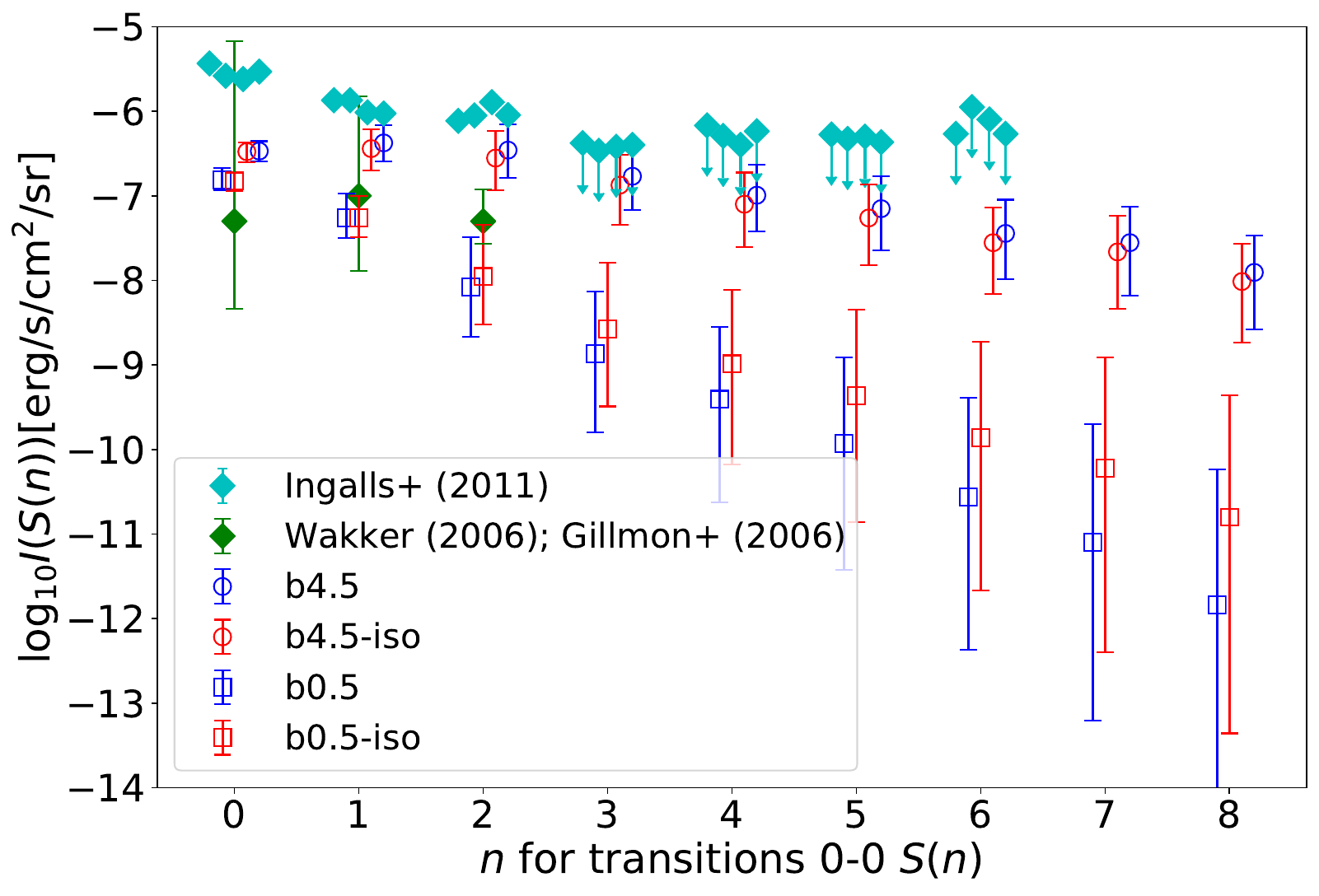}
    \caption{The intensity of each of eight rotational transitions in each of our simulations. The error bars represent 16th and 84th percentiles of the distribution, with the point plotted at the median. The isothermal simulations (after post-processing) look remarkably similar to the non-isothermal simulations. The green points are calculated from {\it FUSE} observations of AGN behind diffuse H$_2$ with $10^{19}<N(\HH)<10^{20}{\rm cm}^{-2}$ \citep{Wakker_2006,Gillmon+Shull+Tumlinson+Danforth_2006}; these sightlines have lower $N_{\rm H}$ than the typical sightlines in our simulations, and are thus expected to differ in intensity by a factor of a few from our simulations. 
    Additional excitation mechanisms appear to be needed to explain the high intensities observed by \citet{Ingalls+Bania+Boulanger+etal_2011}
    for the translucent cloud DCld300.2-16.9, and perhaps also the FUSE observations, being lower column density.
    }
    \label{fig:line_emission}
\end{center}
\end{figure}
%%%%%%%%%%%%%%%%%%%%%%%%%%%%%%%%%%%%%%%%%%%%%%%%%
%
 
Eq.~\ref{eq:sigma2X} has
a species specific temperature, $T_{X}$ for the possibility of ions being out of thermal equilibrium with the neutral species. 
Ion-neutral scattering will heat the ions to a temperature
where heat transfer to and from the ions 
\citep{Draine_1986a} vanishes,
 \begin{equation}
     T_i \approx \TAD + \frac{1}{3k}
     m_nv_d^2
 \end{equation}
 where
 $m_n$ is the mean mass of the neutrals (see Tab.~\ref{tab:parameters}), and $k$ is the Boltzmann constant.
 
 Constructing these velocity profiles for CH$^+$ and $\HH$ in our simulations (Fig.~\ref{fig:chplines}), we see that the CH$^+$ is consistently broader than the $\HH$ velocity profiles. As well, we often see multiple components of CH$^+$, sometimes separated by 10~km/s. For neutral line profiles, while we do see multiple components, as $\sigma_{\rm 3D} \approx 4$~km/s, we do not see component separations of more than a few km/s.

\subsection{$\HH$ rotational line emission, 0-0 $S(n)$}\label{sec:emission}

Figure~\ref{fig:line_emission} shows the intensity of $\HH$ rotational emission lines 0-0 $S(n)$ from $n=0$ through $n=8$ in each of our simulations, computed using the temperature $\TAD$. As is the case with all of our analysis in this paper, we exclude regions where the ion-neutral drift velocity exceeds 5 km/s. The higher field strength simulations (b4.5 and b4.5-iso) exhibit a notably higher level of emission compared to the low field strength simulations (b0.5 and b0.5-iso), especially at higher $n$. Even with the obvious enhancement of the line intensities from the ambipolar diffusion heating, we still do 
not
reach the line intensities seen 
by \citet{Ingalls+Bania+Boulanger+etal_2011}
at four positions on the high-latitude translucent cloud Dcld 300.2-16.9. Other excitation mechanisms  appear to be necessary to explain the \citet{Ingalls+Bania+Boulanger+etal_2011} intensities. However, our simulations \textit{do} largely agree with other data. In Fig.~\ref{fig:line_emission}, we also plot median values of the intensity of 0-0 $S(0)$, $S(1)$, and $S(2)$ for AGN sightlines from \citet{Wakker_2006} and \citet{Gillmon+Shull+Tumlinson+Danforth_2006} towards AGN. As well, the data in Fig.~\ref{fig:chpvsem} from \citet{Lambert+Danks_1986}, \citet{spitzer1974column}, and \citet{FrischJura1980ApJ} show good agreement with both of our simulations.

\newtext{As mentioned in section~\ref{sec:chp}, one point of comparison between observations and our simulations we use is the sum of the intensities of the first three rotational lines 0-0 S(0), 0-0 S(1), and 0-0 S(2). We call this quantity $I(\HH)$. These first three lines contain most of the emission from $\HH$, and so reflect dissipation into $\HH$. 

Observations of $I(\HH)$ plotted in figure~\ref{fig:chpvsem} are computed using known transition rates from the $v=0$, $J=2,3,4$ levels of $\HH$ into the $v=0$, $J=0,1,2$ levels, respectively \citep{Wolniewicz+Simbotin+Dalgarno_1998} together with reported $J$-level column densities. 

For all but one of the points, errors are computed from the errors on the column densities in the source literature. For the one point that is not, the column density for $J=4$ was unavailable. To compute upper and lower bounds on the point, we thus see what the greatest ratio of $N(J=4)/N(J=3)$ is across the observations to compute an upper bound on $N(J=4)$, and similarly to compute a lower bound. We take the (logarithmic) midpoint between the upper and lower bounds to be our value for $N(J=4)$ and compute $S(2)$ from this.}

All of our simulations exhibit a correlation
between $N_{{\rm CH}^+}$ and $I(\HH)$, as is shown in Fig.~\ref{fig:chpvsem}. In our high field strength simulations, b4.5 and b4.5-iso, there is a relatively strong correlation between $I(\HH)$ and $N_{{\rm CH}^+}$, with only $\approx 1$ dex of scatter in $\log_{10}N_{{\rm CH}^+}$ for any given $\log_{10}I(\HH)$. However, at low field strengths, as in b0.5 and b0.5-iso, this correlation is only apparent at higher values of $I(\HH)$, with the correlation disappearing when $I(\HH)\lesssim \text{a few}\times 10^{-8} {\rm erg/s/sr/cm}^2$. As the velocity dispersions $\sigma_{\rm 3D}$ are similar across both field strengths, the explanation for the difference between the joint $N_{{\rm CH}^+}$ vs. $I(\HH)$ histograms must lie in their differing ion-neutral drift velocity distributions. 

In Fig.~\ref{fig:chpvsem}, we also plot six lines of sight towards various stars, with CH$^+$ column densities from \citet{Lambert+Danks_1986} and $I(\HH)$ computed using column densities of the $v=0$, $J=2,3,4$ states of $\HH$ published in \citet{spitzer1974column} and \citet{FrischJura1980ApJ}. These observations agree much better with our simulation data than those from \citet{Ingalls+Bania+Boulanger+etal_2011}. It may be that the \citet{Ingalls+Bania+Boulanger+etal_2011} cloud is in some way atypical compared to these other simulations.

The higher ion-neutral drift velocities present in our simulations b4.5 and b4.5-iso produce higher values of $\TAD$ compared to our low field strength simulations.
Higher values of $\TAD$ mean higher values of $I(\HH)$. Higher values of $\TAD$ also mean higher values of CH$^+$, but as we stated, the drift velocity dependent component of $T_{\rm eff}$ is much more important to CH$^+$ than $\TAD$.

 \subsection{Polarization and the Stokes parameters, $Q$ and $U$}\label{sec:polarization}

%
%%%%%%%%%%%%%%%%%%%%%%%%%%%%%%%%%%%%%%%%%%%%%%%%%
\begin{figure}
\begin{center}
	\includegraphics[width=0.9\columnwidth]{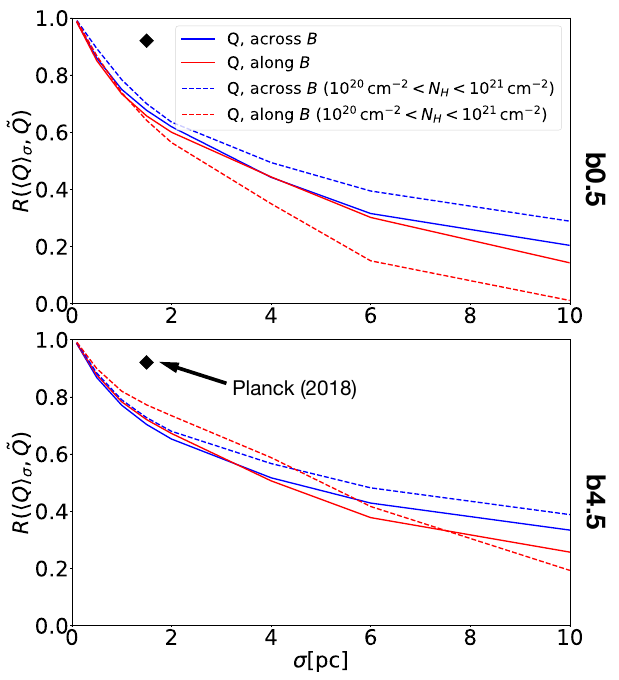}
    \caption{The Pearson correlation coefficient $R$ between $\langle Q \rangle_{\sigma}$ and $\tilde{Q}$ versus $\sigma$ for our simulation b0.5 (top) and b4.5 (bottom) (See Tab.~\ref{tab:sims}). We show these values computed both across (blue) and along (red) the mean field direction. Red and blue dashed lines represent the correlation coefficient when computed over a restricted range of column densities ($N_{\rm H}\in[10^{20},10^{21}]\,{\rm cm}^{-2}$) that are more similar to the \citet{Planck_2018_XII} results. We also show the Planck results for a 40 arcmin beam if the cloud were anywhere between 300 pc away (diamond), where they found $R(\langle Q \rangle_{\sigma}, Q^*) \approx 0.92$ over a wide range of beam sizes and distances. Note that the correlation coefficient decays more quickly for the lower field strength/higher Alfv\'en Mach number.
    }
    \label{fig:pearson}
\end{center}
\end{figure}
%%%%%%%%%%%%%%%%%%%%%%%%%%%%%%%%%%%%%%%%%%%%%%%%%
%

 %
%%%%%%%%%%%%%%%%%%%%%%%% f14 %%%%%%%%%%%%%%%%%%%%%
\begin{figure}
\begin{center}
	\includegraphics[width=1.0\columnwidth]{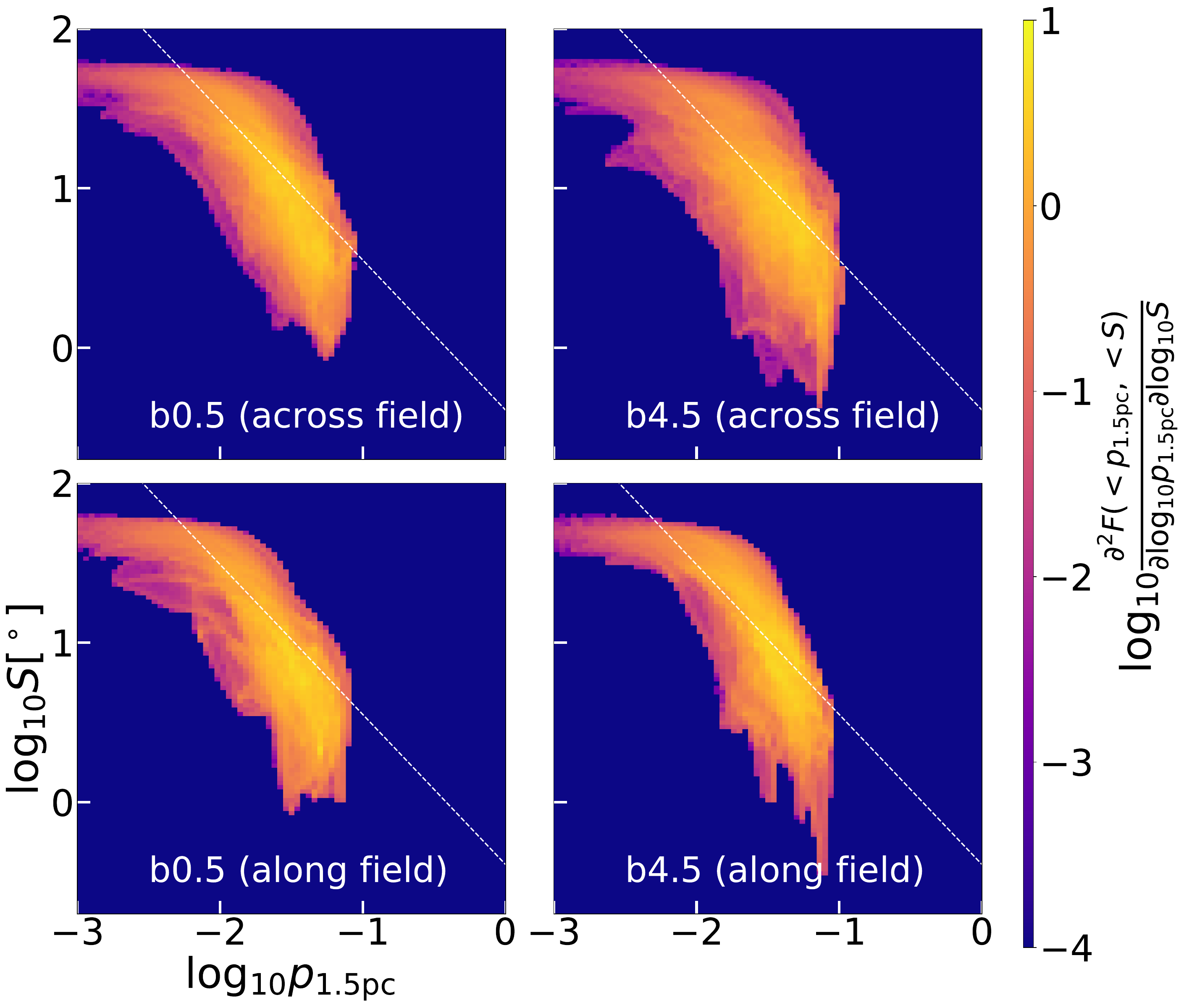}
    \caption{The polarization angle dispersion function $S$ (defined in Sec.~\ref{sec:polarization}) versus the polarization $p$ averaged over a gaussian beam of FWHM = (1.5 pc)$\times8\ln 2$ (see Eq.~\ref{eq:pseudoss}) for each of our simulations b0.5 and b4.5 (see Tab.~\ref{tab:sims}), computed both across the mean field (top) and along it (bottom). We compute $S$ using an annulus of radius of 0.75 pc and a width of 0.75 pc. \citet{planck_int_results_xx} found that (approximately) $\log_{10}S[^\circ] = -0.94\log_{10}p - 0.39$ for polarization $p$ over the Chamaeleon-Musca field (other fields are similar). We find that this relationship (white, dashed) is similar to the one our data exhibits.
    }
    \label{fig:poldisp}
\end{center}
\end{figure}
%%%%%%%%%%%%%%%%%%%%%%%%%%%%%%%%%%%%%%%%%%%%%%%%%
%
 
To compute the polarization of starlight 
and thermal emission due to dust grains along a particular line of sight, 
we compute dimensionless quantities $\tilde{Q}$, $\tilde{U}$, and $\tilde{P}$
 \beqa
     \tilde{Q}_z &=& \frac{\int {\rm d}z \hspace{0.05cm}\rho(B_y^2 - B_x^2)/B^2}{\int {\rm d}z \hspace{0.05cm}\rho}\label{eq:q}\\
     \tilde{U}_z &=& -2\frac{\int{\rm d}z\hspace{0.05cm} \rho B_y B_x /B^2}{\int {\rm d}z \hspace{0.05cm}\rho}\label{eq:u}
\\
     \tilde{P}_z &=& \sqrt{\tilde{U}_z^2+\tilde{Q}_z^2}\label{eq:p},
 \eeqa
 where $B_\alpha$ denotes the component of the magnetic field along direction $\alpha$, and subscripts on each of the $\tilde{Q}$, $\tilde{U}$, and $\tilde{P}$ denote the axis along which we integrate. The mean field direction is along $x$, so the above represent sight-lines \textit{perpendicular to} the mean field. When computing these parameters along the mean field direction, we simply take $(x,y,z)\rightarrow(y,z,x)$.

Starlight from a star behind the cloud would have fractional Stokes parameters
\beq 
(q_\star,u_\star) = N_d C_{{\rm pol},\star} f_{\rm align}
\times
(\tilde{Q}_z,\tilde{U}_z)
\eeq
and fractional polarization
\beq
p_\star = (q_\star^2+u_\star^2)^{1/2} = N_d C_{{\rm pol},\star} f_{\rm align} \times
\tilde P_z ~~,
\eeq
where $C_{{\rm pol},\star}$ is a starlight polarization cross section, and $f_{\rm align}$ measures the degree of alignment
of dust grains with the local magnetic field (see Appendix \ref{app:qup}).

%%%%%%%%%%%%%%%%%%%%% f15 %%%%%%%%%%%%%%%%%%%%%%%%
\begin{figure}
\begin{center}
	\includegraphics[width=1\columnwidth]{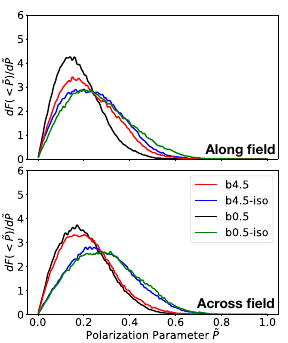}
    \caption{The probability density functions of
    $\tilde{P}$ in our simulations.
    As would be expected, across the field, the higher magnetic field, lower Alfv{\'e}n Mach number simulations have a higher degree of polarization. Viewed along the field, a pattern is less apparent. 
        }
    \label{fig:polarization}
\end{center}
\end{figure}
%%%%%%%%%%%%%%%%%%%%%%%%%%%%%%%%%%%%%%%%%%%%%%%%% 

For thermal emission, the Stokes parameters $Q$ and $U$, and the polarized intensity $P$, 
are directly related to $\tilde{Q}$, $\tilde{U}$, and $\tilde{P}$, but must also be averaged over a beam:
(see Appendix \ref{app:qup}):
\beqa
(\langle Q\rangle_\sigma,
\langle U\rangle_\sigma) &=& B_\nu(T_d) \Cpol f_{\rm align}
\times (
\langle N_d\tilde{Q}\rangle_\sigma,
\langle N_d\tilde{U}\rangle_\sigma)
\\
\label{eq:psigma}
\langle P\rangle_\sigma &=& B_\nu(T_d) \Cpol f_{\rm align}
\times
\left(\langle N_d\tilde{Q}\rangle_\sigma^2 + \langle N_d\tilde{U}\rangle_\sigma^2\right)^{1/2}~,~~~~~~
\eeqa
where $\langle ...\rangle_\sigma$ denotes an average over a beam with beam size parameter $\sigma$, and $B_\nu(T_d)$ is the black-body spectrum of a dust grain at temperature $T_d$.
The fractional polarization is (see Appendix \ref{app:qup})
\beq
p_\sigma \approx \frac{\Cpol}{\Cbar} f_{\rm align} 
\frac{\left(
\langle N_d \tilde{Q}\rangle_\sigma^2 +
\langle N_d \tilde{U}\rangle_\sigma^2
\right)^{1/2}}{\langle N_d\rangle_\sigma}
~~. \label{eq:ppol}
\eeq

\citet{Planck_2018_XII} find that 
$(\langle Q\rangle_\sigma,\langle U\rangle_\sigma)$ and 
$(q_\star,u_\star)$ are highly correlated, with Pearson
correlation coefficient 0.92 for a beam with FWHM=40 arcmin.
\citet{Planck_2018_XII} also find that the ratio $\langle P\rangle_\sigma/p_\star$ changes minimally when the
beam size is varied from 20 arcmin to 80 arcmin.

Fig.\ \ref{fig:pearson} shows the Pearson correlation coefficient $R$ between the beam-averaged $\langle Q\rangle_\sigma$
and the line-of-sight polarization parameter $\tilde{Q}$, as a function of
gaussian beamsize 
\beq
\sigma = 
\frac{\rm FWHM}{\sqrt{8\ln 2}}
= 1.48\pc 
\left(
\frac{\rm FWHM}{40\arcmin}
\right)
\left(\frac{D}{300\pc}\right)
~~,
\eeq
for our b4.5 simulation.
The correlation is shown for random sightlines along the mean field $\hat{\bm x} B_0$, and
across $\hat{\bm x} B_0$.
We see that for this simulation, $R$ drops below $0.9$ for $\sigma > 0.5\pc$.  This contrasts with observations showing $R=0.92$ for $\sigma \approx 1.5\pc$ (for a typical distance $D\approx 300\pc$ to the emitting dust).

In order to study the coherence of the polarization in our simulations, we examine at the polarization angle dispersion function $S$ defined by
\begin{equation}
    S({\bm r}) \equiv \sqrt{\frac{1}{N}\sum_{i=1}^N [\psi({\bm r}) - \psi({\bm r}+{\bm \delta}_i)]^2}
\end{equation}
where the sum runs over all lines of sight in an annulus of radius and width 0.75 pc centered around a position ${\bm r}$, $\psi({\bm r})$ is the polarization angle at ${\bm r}$, and ${\bm \delta}_i$ is an offset that puts ${\bm r}+{\bm \delta}_i$ in the aforementioned annulus \citep{Planck_int_results_xix_2015}. This corresponds roughly to the parameters that \citet{Planck_int_results_xix_2015} used for this function given a beam FWHM of 60 arcmin and an annular radius and width of 30 arcmin if we place our simulation at a distance of $D \approx 300\,{\rm pc}$. For consistency, we also use a gaussian beamsize $\sigma = 1.5$ pc to compute the angles $\psi$. Fig.~\ref{fig:poldisp} shows $S$ versus $p_{1.5\,{\rm pc}}$ for our simulations b0.5 and b4.5. To compute this quantity, we have assumed that $C_{\rm pol}f_{\rm align}/\bar{C} \approx 0.24$. We choose this value because 0.24 is the best estimate for the highest observed level of polarization in \citet{Planck_2018_XII}. Choosing this value for $C_{\rm pol}f_{\rm align}/\bar{C}$ ensures that our computed polarizations are always $\leq 0.24$.

This function has the property that completely random orientations of the polarization yield $S \approx 52^\circ$, explaining the asymptotic behavior at low $p_{1.5\,{\rm pc}}$ in Fig.~\ref{fig:poldisp}. \citet{Planck_int_results_xix_2015,
planck_int_results_xx} found an approximate relationship $S \propto 1/p_{1.5\,{\rm pc}}$ in both their observational data as well as in a colliding flow MHD simulation. We also find this relationship (shown as a white dashed line in Fig.~\ref{fig:poldisp}) 
to approximately hold true for our simulations as well. It is interesting to note that altering the mean field strength/Alfv\'en Mach number seems to have no effect on this relationship. The relationship also seems unaffected by whether or not we are looking along or across the mean magnetic field direction.

Given that our four simulations differ in one of two ways (either field strength or isothermal/non-isothermal), we may examine the effect that each of those variables has on the polarization. Firstly, does including heating and cooling processes throughout the simulation impact the magnetic field geometry? Fig.\ \ref{fig:polarization} shows probability density functions for line-of-sight polarization $\tilde{P}$ in each of our simulations for 512$^2$ sight lines, both across and along the mean field.

We see no marked difference between polarization in simulations run with an isothermal ($\gamma = 1$) equation of state versus simulations run with $\gamma = 5/3$ and heating (cosmic ray, photoelectric emission from dust grains) and cooling ($\HH$, C$^+$, and O) processes included. This suggests that, for the purposes of studying polarization, isothermal MHD turbulence is a reasonable approximation, with heating and cooling processes having only a small effect on  the PDFs of dust polarization along lines of sight.

As might have been expected, the higher field strength simulations exhibit higher values of $\tilde{P}$
(see Fig.~\ref{fig:polarization}), with the difference being most apparent when looking across the mean field. This makes intuitive sense, as the higher field strength simulations are only moderately super-Alfv{\'e}nic, with $\mathcal{M}_A \gtrsim 1$. The magnetic field in those simulations is thus moderately effective at resisting distortions due to the turbulent motions in the cloud. 

When looking down the field, the polarization we see in all simulations will be due only to a random component of the magnetic field. As a result, we do not expect to see polarization levels quite as high looking along the mean field as looking across the field, nor do we expect as large a difference between the polarization levels for different magnetic field strengths. Indeed, this is what we observe in Fig.~\ref{fig:polarization}.

\section{Discussion}\label{sec:discussion}

In our simulations, turbulence and the ambipolar diffusion heating that results from it raise the gas temperature in low density regions to beyond $1000\K$, sufficient to produce CH$^+$. High ion-neutral drift velocities (Eq.~\ref{eq:driftvel}) also enhance the CH$^+$ abundance through increasing the effective reaction temperature (Eq.~\ref{eq:teff}, see also Fig.~\ref{fig:temperatures}). As ion-neutral drift velocities are highest in low-density regions, the CH$^+$ exists predominantly in regions with $n_{\rm H}\sim 1\,{\rm cm}^{-3}$. In some fraction of the volume, these drift velocities exceed 5 km/s, and so have an ambipolar diffusion Reynolds number $R_{\rm AD} \lesssim 1$. As ambipolar diffusion is important to the dynamics in those regions 
but we have not properly included it in the dynamics, we have excluded these regions when calculating CH$^+$ formation and $\HH$ line emission in our models. This is an important caveat of our results, as the details of these regions could be very important to both the CH$^+$ abundance and the $\HH$ rotational line emission (see Fig.~\ref{fig:cutoffs}). 

Our simulations include sight-lines where $N_{\rm H} \lesssim 10^{21}\,{\rm cm}^{-2}$ and $N_{{\rm CH}^+} \gtrsim 10^{13}\,{\rm cm}^{-2}$ (see Fig.~\ref{fig:chpvsnh}.
Such sight-lines are not seen in the data \citep[e.g.][]{weselak2008relation}. 
Because our simulations do not include formation/destruction of $\HH$, even low column density regions are assumed to have a relatively high $\HH$ fraction, and thus are able to form CH$^+$ in our simulation.  In reality, photodissociation will suppress the $\HH$ fraction in low column density regions, where self-shielding is ineffective.
Proper treatment of the variable $\HH$ fraction requires radiative transfer including $\HH$ self-shielding
(V17).

The CH$^+$ column densities found in MML15 lie between the results of our lowest (b0.5 and b0.5-iso) and highest (b4.5 and b4.5-iso) magnetic field strength simulations. In this way, our results are consistent with theirs. We find that ambipolar diffusion can become a significant factor contributing to the heating of the volume, and cannot be neglected energetically. MML15 found that the contribution to the CH$^+$ abundance from the ion-neutral drift velocities to be the dominant effect responsible for producing CH$^+$; we corroborate this finding. The fact that CH$^+$ is predominantly produced by the ion-neutral drift velocities in our model as well as in the MML15 model is the reason that our results qualitatively agree despite the large discrepancy in our $\HH$ cooling functions. Both our paper and theirs rely on the same post-processing method to deduce the ion-neutral drift velocities.

\newtext{As mentioned in section~\ref{sec:intro}, V17's two-phase, colliding flow results contrast with our findings and those  of MML15. V17 found that CH$^+$ was {\it not} primarily produced by high ion-neutral drift velocities in their simulation. As well,} their simulation under-produced CH$^+$ relative to observations, finding column densities $\sim10^{11-12}\,{\rm cm}^{-2}$ rather than $\sim 10^{13}\,{\rm cm}^{-2}$. The reasons for this appear to be three-fold. First, destruction of $\HH$ in low-density regions limits CH$^+$ production by reaction (\ref{eq:react}). Second, a higher ionization fraction means that ion-neutral drift velocities will be lower, and as we find CH$^+$ is very sensitive to changes in the ion-neutral drift velocity, this can have an enormous impact on the overall CH$^+$ abundance.  V17 found overall drift velocities around two orders of magnitude lower than those we find. Third, injecting only warm-phase ISM into the box seems likely to underestimate the $\HH$ fraction in the lower density regions, further decreasing CH$^+$ abundance.

\newtext{We suspect there are also key differences in the way the problem is set up in our simulations versus those of V17. V17 used a 2.5$\mu$G guide field aligned with the direction of their flow. We expect this configuration would naturally lead to less tangling of the field in much of the simulation, in particular in the low density regions where CH$^+$ is likely produced. In contrast, we used a stronger 4.5$\mu$G mean field and drove turbulence isotropically. It is hard to directly compare our results with theirs given the lack of information on their rms magnetic field strength, Alfv\'en Mach number, and velocity dispersion in the CNM phase. It seems guaranteed that a colliding flow simulation of this type would produce results differing from those of a simulation of driven turbulence. It would be interesting to know how well a simulation of the V17 type compares to observations of polarization with the same metrics we have used. To understand this problem, multiphase simulations with realistic driving analogous to stellar winds and supernovae are needed.}

\newtext{Of course, both colliding flow simulations and driven turbulence are artificial in their own ways, and likely to produce discrepant results. It is clear that the main difference between our results and those of V17 is the ion-neutral drift velocity. It is essential that future work accurately capture this physics, or at least explore its effect realistically. The biphasic nature of their simulation also impacts the drift velocity distribution. It would be informative in future work to do a multiphase study with $\HH$ formation and destruction as in V17, but using higher rms magnetic field strengths and a periodic domain driven at large scales the way we have in our study. Recently, \citet{kortgen2020turbulence} showed that turbulence driving in disc galaxies is neither purely compressive nor purely solenoidal.% with a $b\sim 0.5-0.7$. 
Future simulations may employ driving of this type instead of purely solenoidal driving as we have done.}

Our simulations with $B_0=4.5\muG$ produce rotationally-excited $\HH$ at levels approximately consistent with UV absorption measurements of rotationally-excited $\HH$, and the observed correlation between CH$^+$ and rotationally-excited $\HH$ is also reproduced (see Figs.\ \ref{fig:chpvsem} and \ref{fig:line_emission}). However, our models fall short of the strong infrared emission observed from the quiescent translucent cloud DCld\,300.2-16.9 \citep{Ingalls+Bania+Boulanger+etal_2011}.  It appears that some additional process is contributing to the $\HH$ excitation in this cloud.

{\it Planck} observations of polarized emission from dust provide a strong test of MHD simulations.
The fact that the polarization of dust emission in our simulations is so effectively destroyed by averaging over beams a few pc in size (Fig.~\ref{fig:pearson}) is puzzling when compared to the \citet{Planck_int_results_xxii_2015} data. 
Perhaps the mean field $B_0$ should be even stronger than
$B_0=4.5\muG$, thus lowering the Alfv\'en number.  Or perhaps
ambipolar diffusion (or some other field-smoothing mechanism) acts
more strongly than we have assumed in our simulations, in which
ambipolar diffusion has not been treated self-consistently.

In the future, two-fluid simulations that include ambipolar diffusion, full radiative transfer, variable ionization fraction, and time-dependent chemistry will be necessary to fully disentangle all of the separate variables and fully understand the problems we have addressed in this paper. It may not be feasible to model all of these things at once in a volume similar in size to what we have studied here (8000 pc$^3$), and realistically, future simulations will likely capture one or several of these effects at once, unless effective sub-grid models can be developed. 

\section{Summary}\label{sec:summary}
We present MHD simulations of diffuse molecular clouds 20 pc in size using two different mean magnetic field strengths (0.5$\mu$G and 4.5$\mu$G). All simulations have the same 3D velocity dispersion (within 10\%) chosen to remain consistent with the linewidth-size relation for molecular clouds \citep{Solomon+Rivolo+Barrett+Yahil_1987}. 
We compare simulations using an isothermal equation of state with non-isothermal simulations including realistic heating and cooling.
We calculate $\HH$ cooling over a range of temperatures, densities, and $\HH$ fractions, and provide an accurate fitting function for the cooling.
The MHD simulations were post-processed 
in a way similar to MML15 to 
calculate CH$^+$ abundances, and $\HH$ excitation and line emission.  We compare these results to observations
and 
find that 
we can
explain the CH$^+$ abundance (Figs.~\ref{fig:chp},~\ref{fig:chpvsnh}), 
and rotational excitation of $\HH$ in many regions (Figs.\ \ref{fig:chpvsem}, \ref{fig:line_emission}.
However, we fall short of explaining the strong $\HH$ rotational line intensities seen from two translucent clouds
\citep{Ingalls+Bania+Boulanger+etal_2011}. 
The $\HH$ line emission is correlated with the CH$^+$ 
(Fig.~\ref{fig:chpvsem}). 

CH$^+$ appears to be primarily manufactured in low density regions with high ion-neutral drift velocities. These high ion-neutral drift velocities significantly complicate the interpretation of our results because they 
can become 
unphysically large. 
Because of concern about the realism of the high-drift-velocity regions, we calculate CH$^+$ production and $\HH$ line emission only from regions where the drift velocity $v_d < 5$\,km/s,
corresponding to ambipolar diffusion Reynolds number $R_{\rm AD} \lesssim 1$. We examine the effects of this cutoff on total $\HH$ rotational line emission and CH$^+$ abundance in figure~\ref{fig:cutoffs}. 

We also construct synthetic line-of-sight velocity distributions for CH$^+$ and $\HH$ molecules (Fig.~\ref{fig:chplines}). The CH$^+$ profiles tend to be broader than the neutral line profiles, especially in simulations with a higher mean magnetic field strength. 

We compute the polarization of starlight and thermal emission from dust grains and in our MHD simulations, and compare our results to those of \citet{Planck_2018_XII}. The polarization statistics are largely unaffected by which method was used for the hydrodynamics (including heating and cooling processes vs. using an isothermal equation of state).  The polarization does depend on field strength/Alfv\'{e}n Mach number, with higher field strength (lower Alfv\'{e}n Mach number) corresponding to higher levels of polarization (Fig.~\ref{fig:polarization}). 

We examine the effect of averaging the Stokes $Q$ and $U$ over a gaussian beam of varying width and find our simulations to be marginally inconsistent with the findings of \citet{Planck_2018_XII}. 

The correlation between beam-averaged polarization of dust thermal emission and  
polarization of starlight along a sightline is too low in our simulations. We suspect that this is due to our Alfv\'{e}n Mach number in our simulations ($\mathcal{M}_A\sim 1.2-1.4$) being higher than that in nature (perhaps $\mathcal{M}_A\lesssim 1$) on these scales. 
We speculate that perhaps field-smoothing by ambipolar diffusion in nature is more effective than in our simulations, which do not treat ambipolar diffusion self-consistently.

We conclude that the Alfv\'{e}n Mach number in interstellar clouds is likely smaller than in our simulations, likely $\mathcal{M}_A \lesssim 1$. 
Going beyond the work we have done here will require two-fluid simulations to self-consistently model the effects of ambipolar diffusion in multiphase regions. 

\section*{Acknowledgements}

We thank 
Vincent Guillet, Brandon Hensley, Chang-goo Kim, Matthew Kunz, Chris McKee, Eve Ostriker, and Dan Welty
for many valuable discussions. We also thank Ivanna Escala for great advice on data visualization. As well, we thank the anonymous reviewer for a careful reading of our manuscript and many helpful comments.
This research was supported in part by 
NSF grants AST-1408723 and AST-1908123.

KT was supported by the Japan Society for the Promotion of Science (JSPS) KAKENHI Grant Numbers 16H05998, 16K13786, 17KK0091, 18H05440.

\section*{Data availability}
The data presented in this paper was generated on the Princeton computing cluster `Perseus' and will be made freely available upon request to the corresponding author.

%%%%%%%%%%%%%%%%%%%%%%%%%%%%%%%%%%%%%%%%%%%%%%%%%%



%\bibliographystyle{mnras}

\bibliographystyle{mnras}
\bibliography{ms_reduced}

%%%%%%%%%%%%%%%%%%%%%%%%%%%%%%%%%%%%%%%%%%%%%%%%%%

%%%%%%%%%%%%%%%%% APPENDICES %%%%%%%%%%%%%%%%%%%%%

\appendix

\section{Numerical Convergence}\label{app:cutoff}

%
%%%%%%%%%%%%%%%%%%%%%%%%%%%%%%%%%%%%%%%%%%%%%%%%%
\begin{figure*}
\begin{center}
	\includegraphics[width=2.0\columnwidth]{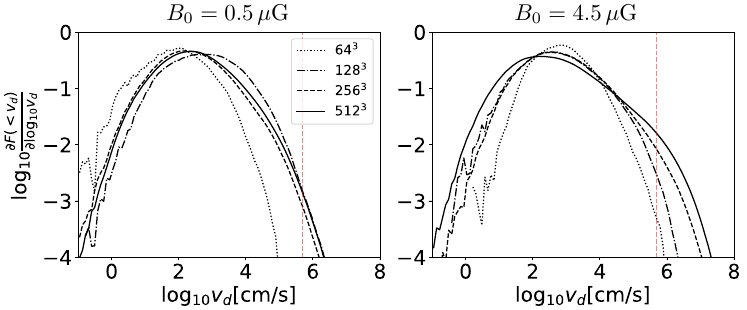}
    \caption{Probability density functions of the ion-neutral drift velocity at various resolutions for each of our chosen mean field strengths, 0.5 $\mu$G and 4.5 $\mu$G. The vertical dashed red line represents our 5 km/s cut that we employ throughout the paper. While it does not appear that our results fully converge, we would not necessarily expect them to: turbulence is an inherently unresolved problem in the absence of physical dissipation.
    }
    \label{fig:driftvels}
\end{center}
\end{figure*}
%%%%%%%%%%%%%%%%%%%%%%%%%%%%%%%%%%%%%%%%%%%%%%%%%
%
One result of our study has been that the 
%main contribution to the CH$^+$ abundance and $\HH$ rotational line luminosities is the ion-neutral drift velocity. 
the ion-neutral drift velocity $v_d$ plays a critical role in CH$^+$ formation and excitation of $\HH$ rotational line emisssion.
Here we examine the sensitivity of $v_d$ to resolution for our simulations with heating and cooling included throughout and an adiabatic $\gamma = 5/3$. In figure~\ref{fig:driftvels}, we show the probability density functions of the ion-neutral drift velocity at four different resolutions, from $64^3$ through $512^3$, and at two separate mean magnetic field strengths, 0.5~$\mu$G and 4.5~$\mu$G. It is worth noting that the highest end of this distribution will never converge in a turbulent domain without ambipolar diffusion being modeled explicitly. This provides additional reason for us to implement a cut in the ion-neutral drift velocities that we consider in our analysis. 

\begin{figure}
\begin{center}
	\includegraphics[width=1.0\columnwidth]{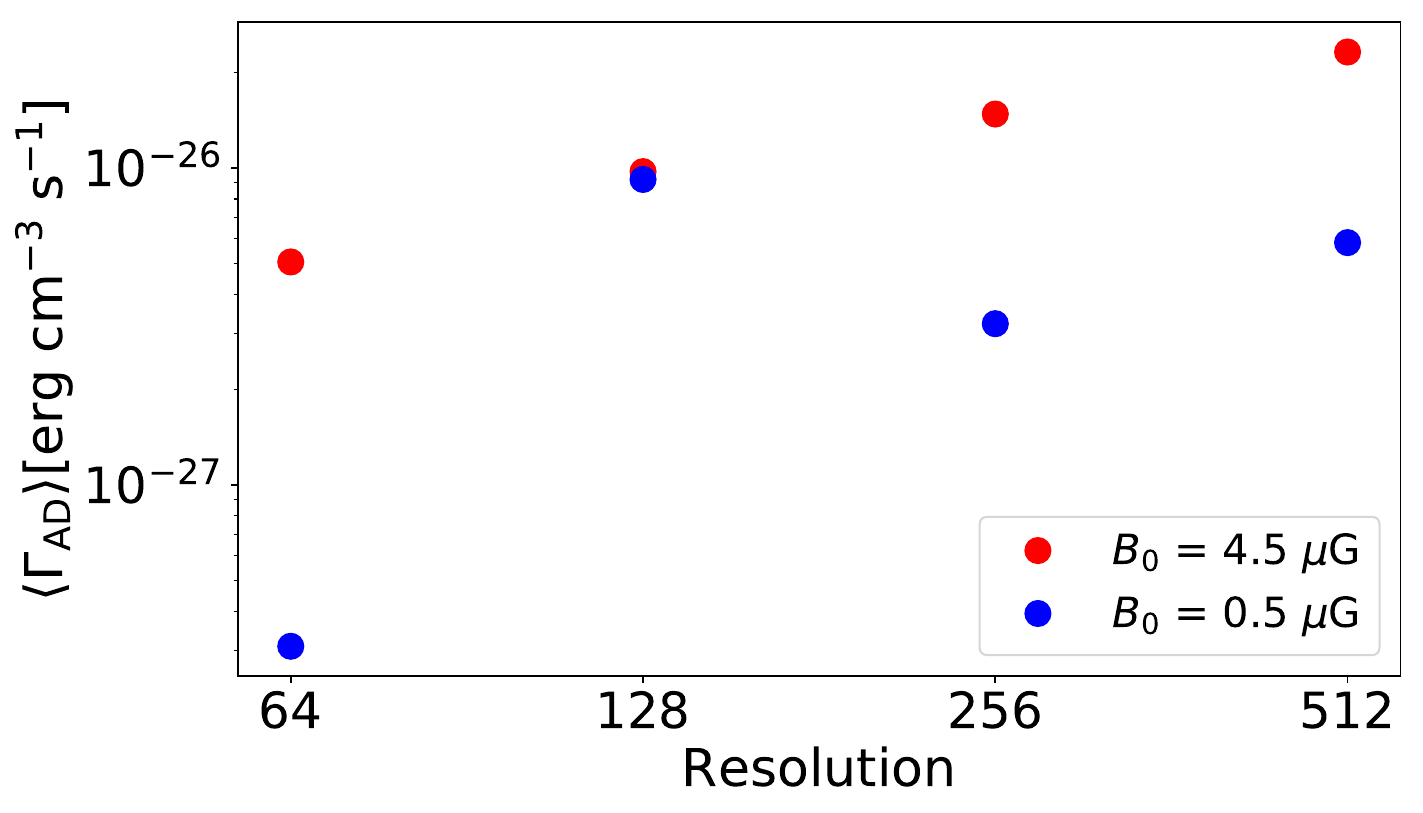}
    \caption{Volume averaged ambipolar diffusion heating calculated for only those regions where the drift velocity $v_d < 5$ km/s. Points in blue are for the low (0.5 $\mu$G) field strength simulations, while points in red are for the high (4.5 $\mu$G) field strengths.
    }
    \label{fig:adheat}
\end{center}
\end{figure}

Figure~\ref{fig:adheat} shows the mean ambipolar diffusion heating $\langle \Gamma_{\rm AD}\rangle$ computed only for cells with ion-neutral drift velocities below 5~km/s. While our results do seem to have some resolution dependence, we would expect this. Only if we were to introduce an explicit physical dissipation mechanism would we expect a well-converged result.

%
%%%%%%%%%%%%%%%%%%%%%%%%%%%%%%%%%%%%%%%%%%%%%%%%%
%%%%%%%%%%%%%%%%%%%%%%%%%%%%%%%%%%%%%%%%%%%%%%%%%
%

\section{Calculating $\HH$ level populations and emission}\label{app:hh} %formerly \label{app:rates}
%\section{Collisional Rate Coefficients}\label{app:rates}

\subsection{Radiative Processes}

We seek to calculate the populations of the $(v,J)$ levels of the
$\HH$ electronic ground state $X^1\Sigma_g^+$.
Einstein $A$ coefficients for quadrupole transitions between the
different $(v,J)$ levels are taken from
\citet{Turner+Kirby-Docken+Dalgarno_1977}.\footnote{%
   For all important transitions, these are in agreement with
   the more recent values from
   \citet{Wolniewicz+Simbotin+Dalgarno_1998}.}

Our code allows for photoexcitation out of the ground electronic state to
the $B^1\Sigma_u^+$, $C^1\Pi_u^+$, and $C^1\Pi_u^-$ states,
followed by either dissociation or return to bound $(v,J)$
levels of the ground electronic state.  We use energy levels,
Einstein $A$ coefficients, and dissociation probabilities from
\citet{Abgrall+Roueff_1989} and
\citet{Abgrall+Roueff+Launay+etal_1993a,
       Abgrall+Roueff+Launay+etal_1993b}.
However, the results presented here assume ultraviolet pumping 
to be weak enough so that $\HH$
excitation is dominated by collisional processes.

\subsection{Collisional Rate Coefficients}

We include the effects of collisions with $\Ha$, $\He$, $e^-$, and
$\HH$ on the rovibrational excitation and deexcitation of $\HH$.
Excitation rates are obtained from deexcitation rates using detailed balance:
\beq
k_{\ell\rightarrow u} 
= 
\frac{g_n(J_u)}{g_n(J_\ell)} \left(\frac{2J_u+1}{2J_\ell+1}\right) 
e^{-(E_u-E_\ell)/kT}\,
k_{u \rightarrow \ell} ~~,
\eeq
where $E_u>E_\ell$, 
and $g_n(J)=1,3$ for $J=$even, odd.

\subsection{Statistical Equilibrium}

Let $\nH\equiv n(\Ha)+2n(\HH)+n(\Ha^+)$ be the number density of H nucleons.
We consider the $299$ 
bound rotation-vibration levels of $\HH$ with rotational
quantum number $J\leq29$, and assign level index $i$ in order of increasing
energy $E_i$, with $i=0$ corresponding to $(v\!=\!0,J\!=\!0)$ and $i=298$ to
$(v\!=\!14,J\!=\!3)$.
Let
\beq 
x_i \equiv \frac{2n(\HH(v_i,J_i))}{\nH} \hspace*{1.0cm},\hspace*{1.0cm}i=0,N-1
\eeq
be the fraction of the H nucleons
in $\HH(v_i,J_i)$, and
let
\beq
x_N \equiv \frac{n(\Ha)}{\nH}
\eeq
be the fraction of the $\Ha$ nucleons in atomic H.
We neglect the small fraction of H in molecules other than $\HH$.
The ionized fraction is then $n(\Ha^+)/\nH=1-\sum_{j=0}^N x_j$.

For $f\neq i$ we define a transition matrix
\beq
R_{if} = k_{if}^{\Ha}n(\Ha) + k_{if}^{\He}n(\He)
+ k_{if}^{\HH}n(\HH) + A_{if} + \zeta_{if} + \Phi_{if}
\eeq
where, for $0\leq i<N$
\beqa
k_{if}^X \hspace{-0.3cm}&~=~& \hspace{-0.3cm}{\rm rate~ coefficient~ for~}
X + \Ha(v_i,J_i) \rightarrow X + \HH(v_f,J_f)\nonumber \\
&&{\rm ~for~} f<N
\\
k_{iN}^X \hspace{-0.3cm}&=& \hspace{-0.3cm}{\rm rate~ coefficient~ for~}
X + \Ha(v_i,J_i) \rightarrow X + 2\Ha,
\\
A_{if}\hspace{-0.3cm} &=&\hspace{-0.3cm} {\rm spontaneous~ decay~ rate~} (f<i<N),
\\
\zeta_{if}\hspace{-0.3cm} &=&\hspace{-0.3cm}  i\rightarrow f {\rm ~transition~ rate~
due~ to~ UV~ pumping~},\nonumber\\
&&(i<N, f<N),
\\
\zeta_{iN} \hspace{-0.3cm}&=&\hspace{-0.3cm} {\rm photodissociation~ rate~}
(\HH(v_i,J_i) + h\nu \rightarrow 2\Ha)
\\
\Phi_{if}\hspace{-0.3cm}&=&\hspace{-0.3cm}0 {\rm ~for~} i<N,
\\
\Phi_{Nf}\hspace{-0.3cm}&=&\hspace{-0.3cm}R_{gr}\nH \phi_f {\rm ~for~} f<N ~~,
\eeqa
where
$R_{gr}\nH n(\Ha)$ is the rate per volume of $\HH$ formation on grains,
and $\phi_f$ is the fraction of newly-formed $\HH$ in rotation-vibration state
$(v_f,J_f)$. 

For convenience, we define the diagonal elements
\beq
R_{ii} = -\sum_{f\neq i}R_{if} ~~.
\eeq
Then we have
\beq
\frac{d}{dt} x_f = \sum_{i=0}^N R_{if} x_i ~~~~{\rm for~}f=0, ..., N ~~.
\eeq

\subsection{Pseudo Steady State}

The true steady state solution with $dx_i/dt=0$ for $i=0,...,N$ is one where
$\HH$ dissociation is balanced by $\HH$ formation.  
The time scale for approaching this steady state, 
$\tau = (2\nH R_{gr} + 2\sum_i \zeta_{iN}x_i)^{-1}$, 
is generally long compared to the timescales for redistribution over the
vibration-rotation states.
In fact, $\HH$ formation and dissociation will often not be balanced in
interstellar molecular gas, 
particularly hot gas that may be cooling rapidly after being shock-heated.

%%%%%%%%%%%%%%%%%%%%%%%%%%%%%%%%%%%%%%%%%%%%%%%%%%%%%%%%%%%%%%%%%%%%%%%
%\begin{figure*}
%\begin{center}
%\includegraphics[width=0.5\textwidth]%l b r t
%               {figures/fclma.pdf}
%\includegraphics[width=0.5\textwidth]%l b r t
%               {figures/fclmm.pdf}
%\includegraphics[width=8.0cm,angle=0,
%                 clip=true,trim=0.5cm 4.5cm 0.5cm 0.5cm]%l b r t
%               {figures/fclma.pdf}
%\hspace*{-0.2cm}\includegraphics[width=8.0cm,angle=0,
%                clip=true,trim=0.5cm 4.5cm 0.5cm 0.5cm]%l b r t
%                {figures/fclmm.pdf}
%\label{fig:CCM19}
%\caption{
%\footnotesize
%Cooling per $\HH$ for densities $\nH=1$, $10$, $10^2$, $10^3$, $10^4$, and
%$10^5\cm^{-3}$.
%(a) Gas which is 99\% atomic and 1\% $\HH$.
%(b) Gas which is 99\% $\HH$ and 1\% atomic.
%Solid curves: present calculations.
%Dashed curves: fitting function from
%\citet[][hereafter CCM19]{Coppola+Lique+Mazzia+etal_2019}.
%For each case, 
%$\sigma=\langle\left[\ln(\Lambda/\Lambda_{\rm fit})\right]^2\rangle^{1/2}$ 
%measures the quality of the CCM19 fitting function 
%over the range $100\leq T\leq 4000\K$.
%}
%\end{center}
%\end{figure*}

Collisional deexcitation of levels that have been populated by
UV pumping and injection of newly-formed $\HH$ in excited states
can act as a heat source, confusing the calculation of collisional
cooling.
%\omittext{We will assume the $\Ha^+$ fraction is small and constant.}
To remove the effects of UV pumping and isolate the 
collisional cooling, 
we treat the atomic fraction $x_N \equiv n(\Ha)/\nH$ as a parameter,
artificially suppress the rate of $\HH$ formation by a factor 
$\lambda_{\rm form}\ll 1$,
and find the solution to the set of equations
\beq \label{eq:pseudoss}
0 = \sum_{i=0}^{N} R_{if}^\prime
x_i ~~~~~{\rm for~}f=0, ..., N~~,
\eeq
where
\beqa
R_{if}^\prime &= 
k_{if}^{\rm H} n(\Ha) +
k_{if}^{{\rm H}_2} n(\HH) +
k_{if}^{\rm He} n(\He) +
A_{if} +
\lambda_{\rm uv}\zeta_{if}\\
&~~~~~{\rm for}~ i\neq f ~,~ i<N~,~f<N \nonumber
\\
R_{iN}^\prime &=
\lambda_{\rm uv}
\left[
k_{iN}^\Ha n(\Ha) + k_{iN}^\He n(\He) + k_{iN}^{\HH} n(\HH) + \zeta_{iN}
\right]~~,
\\&~~~~~{\rm for}~i<N \nonumber
\\
R_{Ni}^\prime &\equiv \lambda_{\rm form}R_{gr}\nH \phi_i \\&~~~~~{\rm for}~i<N \nonumber
\\
R_{ii}^\prime &\equiv -\sum_{f\neq i} R_{if}^\prime
~~~,
\eeqa
where the factor $\lambda_{\rm uv}$ modifies all of the dissociation rates to
ensure that the steady-state solution to (\ref{eq:pseudoss})
has the desired atomic fraction $x_N$:
\beq
\lambda_{\rm uv} \equiv
\lambda_{\rm form} \frac{2 R_{\rm gr}\nH x_N}{\sum_{i=0}^{N-1}x_i R_{iN}}
~.
\eeq
The rates for UV pumping are also suppressed by the factor $\lambda_{\rm uv}$.
Because we take $\lambda_{\rm form}\ll 1$, the H$_2$ level populations are determined only by collisional processes and spontaneous radiative decay.

\subsection{$\HH(v,J) + \Ha$}

\citet{Lique_2015} has calculated collisional deexcitation rate coefficients for
$\HH(v_u,J_u)+\Ha\rightarrow \HH(v_\ell,J_\ell)+\Ha$ for the
63 rotation-vibration excited states with
$E(v_u,J_u)/hc < 15240 \cm^{-1}$ ($E/k < 21930\K$), for temperatures
$100\K\leq T \leq 5000\K$.
This includes levels up to $(v,J)=(0,17)$, $(1,14)$, $(2,11)$, and $(3,8)$.
We use the \citet{Lique_2015} deexcitation rates for deexcitation from these
levels.
For deexcitation from levels $E(v_u,J_u)/hc > 15240\cm^{-1}$, we use
the \citet{Lique_2015} rates for the same $J_u$, $J_\ell$, and $v_u-v_\ell$
if available; otherwise we use rates for the same $\Delta v$ and the highest
$J_u$ considered by \citet{Lique_2015}.
For $100\K <T<5000\K$ we interpolate in the rates provided by
\citet{Lique_2015}.
For $T<100\K$ or $T>5000\K$ we assume the deexcitation rates to scale as 
$k \propto \sqrt{T}$.

\subsection{$\HH(v,J) + \HH$}

For collisional deexcitation
$\HH(v_u,J_u)+\HH \rightarrow \HH(v_\ell,J_\ell)+\HH$ we assume that one of the
colliding $\HH$ molecules does not change state; 
this is clearly incorrect, but a full set of state-to-state rate 
coefficients for $\HH+\HH$ collisions
is not yet available.
We use the inelastic
cross sections calculated by
\citet{LeBourlot+PineaudesForets+Flower_1999} increased by a factor of
3 to obtain vibrational relaxations rates in agreement with
the experimental results of \citet{Dove+Teitelbaum_1974} at
$T=1000$, $2000$, and $3000\K$.

For $\HH(v_u,J_u)+\He\rightarrow\HH(v_\ell,J_\ell)+\He$ we use simple analytic
fits to the rates calculated by
\citet{LeBourlot+PineaudesForets+Flower_1999}.

\subsection{$\HH(v,J) + \He$}

For collisional deexcitation
$\HH(v_u,J_u) + \He \rightarrow \HH(v_\ell,J_\ell)+\He$ we use the analytic
functions provided by \citet{LeBourlot+PineaudesForets+Flower_1999} to fit their
quantum-mechanical results.
The results of \citet{Balakrishnan+Vieira+Babb+etal_1999} 
appear to be in good agreement with the 
\citet{LeBourlot+PineaudesForets+Flower_1999} rates for the
$|\Delta J|\leq 4$ transitions that dominate the inelastic collisions.

\subsection{$\HH(v,J) + e^-$}

We consider only $\Delta J=0$ and $\pm2$.

For $\HH(v_u,J_u)+e^-\rightarrow \HH(v_u,J_u-2)+e^-$ we use
rates obtained from the $J=0\rightarrow2$ experimental data of
\citet{Crompton+Gibson+McIntosh_1969} and the $J=1\rightarrow3$ data of
\citet{Linder+Schmidt_1971}:
\begin{align}
k[(v_u,J_u)&\rightarrow (v_u,J_u-2)]\nonumber
= \\ &2.4\times10^{-11}\frac{J_u(2J_u-3)}{2J_u-1}
\frac{\left(1+ kT/\Delta E\right)^{3/2}}{1+ 10^{-3}T_3^2} \cm^3\s^{-1}
\end{align}
For $\Delta v=-1$ we take
\begin{align}
k[(v_u,J_u)&\rightarrow (v_u-1,J_\ell)] \nonumber
= \\ &1.2\times10^{-10} v_u f(J_u,J_\ell) 
\left(\frac{T_3+0.8T_3^{1.2}}{1+0.005T_3^2}\right) 
\cm^3\s^{-1}~~,
\end{align}
with
\begin{align}
f(J_u,J_\ell) &~\equiv~& \frac{2J_\ell+1}{4J_u+6} ~~~~{\rm for~}J_u\leq 1
\\
&\equiv& \frac{2J_\ell+1}{6J_u+3} ~~~~{\rm for~}J_u\geq 2 ~~.
\end{align}
For $\Delta v=-2$ and $\Delta v=-3$
we take
\beqa
&k[(v_u,J_u)\rightarrow (v_u-2,J_\ell)]
=\nonumber \\
&3\times10^{-12} v_u f(J_u,J_\ell) \frac{T_3^{1.6}}{1+0.003T_3^2}
\cm^3\s^{-1}
\\
&k[(v_u,J_u)\rightarrow (v_u-3,J_\ell)]
=\nonumber\\
&1.2\times10^{-12} v_u f(J_u,J_\ell) \frac{T_3^{1.5}}{1+0.018T_3^2}
\cm^3\s^{-1} ~~.
\eeqa
For $\Delta v=-4,-5,-6$ we take
\beq
k[(v_u,J_u)\rightarrow (v_\ell,J_\ell)]
=\nonumber
2\times10^{-13} v_u f(J_u,J_\ell) \frac{T_3^{1.5}}{1+0.01T_3^2}
\cm^3\s^{-1}
\eeq

\section{Polarization by Aligned Dust Grains}
\label{app:qup}
Suppose the dust grains to be oblate spheroids, with cross sections
$C_a$ and $C_b$ for $\bE$ parallel and perpendicular to the symmetry axis 
$\bahat$.
Consider directions $\bxhat$ and $\byhat$ in the plane of the sky.
In the Rayleigh limit $\lambda \gg a_{\rm eff}$, 
a dust grain will have cross sections
\beqa \label{eq:MPFA1}
C_x &~=~& C_a (\bahat\cdot\bxhat)^2 + C_b [1-(\bahat\cdot\bxhat)^2]
\\ \label{eq:MPFA2}
C_y &=& C_a (\bahat\cdot\byhat)^2 + C_b [1-(\bahat\cdot\byhat)^2]
\eeqa
for radiation with $\bE$ in the $\bxhat$ and $\byhat$ directions, 
respectively.
Let $\falign$ measure the alignment of grain axes 
$\bahat$ with the local magnetic field direction $\bbhat\equiv\bB/B$:
\beqa
\falign &~\equiv~& 
\frac{3}{2}\left[\langle (\bahat\cdot\bbhat)^2\rangle - \frac{1}{3}\right]
~~~.
\eeqa
Randomly-oriented grains have $\langle (\bahat\cdot\bbhat)^2\rangle=1/3$ and
$\falign=0$; perfectly-oriented grains have $\falign=1$.
We assume $\falign$ to be independent of position.
Define
\beqa
\beta_x(x,y)&~\equiv~& \frac{\int dz\, \rho\, (\bbhat\cdot\bxhat)^2}{\int dz\, \rho} 
\\
\beta_y(x,y)&\equiv& \frac{\int dz\, \rho\, (\bbhat\cdot\byhat)^2}{\int dz\, \rho} 
\\
\beta_{xy}(x,y) &\equiv& 2\,\frac{\int dz\, \rho\, 
(\bbhat\cdot\bxhat)(\bbhat\cdot\byhat)}{\int dz\, \rho}
\\
\tilde{Q}(x,y) &~\equiv~& \beta_x-\beta_y
\\
\tilde{U}(x,y) &~\equiv~& -\beta_{xy}
\\
\tilde{P}(x,y) &\equiv& \left(\tilde{Q}^2+\tilde{U}^2\right)^{1/2}
~~~.
\eeqa

Suppose the grains have $\bahat\cdot\bbhat=\cos\theta$, with $\bahat$ precessing around $\bbhat$.
Averaging over the precession and along the sightline:
\beqa
\langle(\bahat\cdot\bxhat)^2\rangle 
&~=& \falign \beta_x +
\frac{1}{3}\left(1-\falign\right)
\\
\langle (\bahat\cdot\byhat)^2\rangle &=& \falign \beta_y +
\frac{1}{3}\left(1-\falign\right)
~~~.
\eeqa
Define
\beqa
\Cbar &~\equiv~& \frac{2C_b+C_a}{3}
\\
\Cpol &\equiv& \frac{C_b-C_a}{2}
~~~.
\eeqa
In the Rayleigh limit, an optically-thin sightline has 
emitted intensity
\beqa
I &=& \frac{1}{2} 
N_d B_\nu(T_d)
\left[\langle C_x\rangle + \langle C_y\rangle\right]
\\
&=& 
N_d B_\nu(T_d)
\left[\Cbar + \falign \Cpol\left(\frac{2}{3}-\beta_x-\beta_y\right)\right]
~~~.
\eeqa
The Stokes $Q$ and $U$, and polarized intensity $P$ are
\beqa \nonumber
Q &~=~& \frac{1}{2} N_d B_\nu(T_d) (\langle C_x\rangle -\langle C_y\rangle)
\\ \nonumber
&=& N_d B_\nu(T_d) \Cpol \falign (\beta_y-\beta_x)
\\
&=& N_d B_\nu(T_d) \Cpol \falign \tilde{Q}
\\ \nonumber
U &=& -N_d B_\nu(T_d) \Cpol \falign \beta_{xy}
\\
&=& N_d B_\nu(T_d) \Cpol \falign \tilde{U}
\\
P &=& (Q^2+U^2)^{1/2}
~~.
\eeqa
Let $\langle ...\rangle_\beam$ denote a beam average.  The polarized intensity and total intensity are 
\beqa
\langle P\rangle_\beam &=& B_\nu(T_d)\Cpol \falign
\left( 
\langle N_d \tilde{Q}\rangle_\beam^2 + 
\langle N_d \tilde{U}\rangle_\beam^2
\right)^{1/2}
\\ \nonumber
\langle I\rangle_\beam &=& B_\nu(T_d)\bar{C} \langle N_d\rangle_\beam \times
\\ \nonumber
&&
\left[1  + 
\falign \frac{\Cpol}{\bar{C}}\left(\frac{2}{3}
- \frac{\langle N_d \beta_x\rangle_\beam}{\langle N_d\rangle_\beam} 
- \frac{\langle N_d \beta_y\rangle_\beam}{\langle N_d\rangle_\beam}
\right)
\right]
\\
&\approx&
B_\nu(T_d)\bar{C} \langle N_d\rangle_\beam
\eeqa
The beam-averaged fractional polarization is
\beq
p =~\frac{\langle P\rangle_\beam}{\langle I\rangle_\beam}
\approx
\frac{\Cpol}{\bar{C}}\falign
\frac{
\left( 
\langle N_d \tilde{Q}\rangle_\beam^2 + 
\langle N_d \tilde{U}\rangle_\beam^2
\right)^{1/2}}{\langle N_d\rangle_\beam}
~~~.
\eeq
Aligned grains polarize starlight.
Let $C_{a,\star}$ and $C_{b,\star}$ be extinction cross sections for starlight
polarized parallel or perpendicular to $\bahat$.  
The ``modified picket fence approximation'' (MPFA) consists of using 
Eq.\ (\ref{eq:MPFA1},\ref{eq:MPFA2}) 
even at wavelengths that are comparable to the grain size.
Draine \& Hensley (2020, in prep.) show that the 
MPFA is sufficiently accurate to use in modeling starlight polarization. 
The difference in extinction cross section for 
radiation polarized perpendicular or parallel to the projection
of $\bB$ on the plane of the sky is then
\beq
C_{{\rm ext}\star,\parallel}- C_{{\rm ext}\star,\perp}
= \left[C_{b,\star}-C_{a,\star}\right]
f_{\rm align}[(\bbhat\cdot\bxhat)^2+(\bbhat\cdot\byhat)^2]
\eeq
and initially unpolarized radiation develops a polarization characterized by
\beqa
(q_\star,u_\star) &\approx& N_d 
C_{{\rm pol}\star} \falign
(\tilde{Q},\tilde{U})
\\
C_{\rm pol,\star} &\equiv& \frac{C_{b,\star}-C_{a,\star}}{2}
~~.
\eeqa
The fractional polarization of the starlight is
\beq
p_\star = \left( q_\star^2 + u_\star^2 \right)^{1/2}
~~.
\eeq

%%%%%%%%%%%%%%%%%%%%%%%%%%%%%%%%%%%%%%%%%%%%%%%%%%

% Don't change these lines
\bsp	% typesetting comment
%\label{lastpage}
\end{document}